\documentclass[a4paper,fleqn]{cas-dc}
\usepackage{textcomp}
\usepackage[numbers]{natbib}
\usepackage{flushend}
\def\tsc#1{\csdef{#1}{\textsc{\lowercase{#1}}\xspace}}
\tsc{WGM}
\tsc{QE}
\tsc{EP}
\tsc{PMS}
\tsc{BEC}
\tsc{DE}

\newtheorem{thm}{Theorem}

\newtheorem{lema}{Lemma}

\newproof{proof}{Proof}

\begin{document}
\let\WriteBookmarks\relax
\def\floatpagepagefraction{1}
\def\textpagefraction{.001}

\shorttitle{Forecasting of Cholera outbreak using epidemic-informed machine learning}

\shortauthors{Ghosh et al. (2025)}

\title [mode = title]{Developing cholera outbreak forecasting through qualitative dynamics: Insights into Malawi case study}                      


%
\author[1]{Adrita Ghosh}





\credit{Conceptualization, Methodology, Software, Formal analysis, Investigation, Visualization, Writing -- original draft and editing}

\address[1]{Department of Mathematics, Indian Institute of Engineering Science and Technology, Shibpur,
West Bengal 711103, India} 

\author[2]{Parthasakha Das}

\address[2]{Department of Mathematics, Rajiv Gandhi National Institute of Youth Development, Sriperumbudur, Tamil Nadu 602105, India}
\credit{Conceptualization, Methodology, Software, Formal analysis, Investigation, Visualization, Writing - original draft \& editing}
\author[3,4]{Tanujit Chakraborty}
\credit{ Conceptualization, Methodology, Software, Formal analysis, Investigation, Visualization,  Writing -- original draft and editing}
\address[3]{SAFIR, Sorbonne University Abu Dhabi, Abu Dhabi, United Arab Emirates}
\address[4]{Sorbonne Centre for Artificial Intelligence, Sorbonne University, Paris 75006, France}


\author%
[1]
{Pritha Das}
\credit{Supervision, Validation, Visualization,  Writing -- original draft and editing}


\author%
[5]
{Dibakar Ghosh}
\cormark[1]
\credit{Supervision, Validation, Visualization,  Writing -- original draft and editing}
\address[5]{Physics and Applied Mathematics Unit, Indian Statistical Institute, 203 B. T. Road, Kolkata 700108, India}

\cortext[cor1]{Corresponding author: dibakar@isical.ac.in (Dibakar Ghosh)}



\begin{abstract}
Cholera, an acute diarrheal disease, is a serious concern in developing and underdeveloped areas. A qualitative understanding of cholera epidemics aims to foresee transmission patterns based on reported data and mechanistic models. The mechanistic model is a crucial tool for capturing the dynamics of disease transmission and population spread. However, using real-time cholera cases is essential for forecasting the transmission trend. This prospective study seeks to furnish insights into transmission trends through qualitative dynamics followed by machine learning-based forecasting. The Monte Carlo Markov Chain approach is employed to calibrate the proposed mechanistic model. We identify critical parameters that illustrate the disease's dynamics using partial rank correlation coefficient-based sensitivity analysis. The basic reproduction number as a crucial threshold measures asymptotic dynamics. Furthermore, forward bifurcation directs the stability of the infection state, and Hopf bifurcation suggests that trends in transmission may become unpredictable as societal disinfection rates rise. Further, we develop epidemic-informed machine learning models by incorporating mechanistic cholera dynamics into autoregressive integrated moving averages and autoregressive neural networks. We forecast short-term future cholera cases in Malawi by implementing the proposed epidemic-informed machine learning models to support this. We assert that integrating temporal dynamics into the machine learning models can enhance the capabilities of cholera forecasting models. The execution of this mechanism can significantly influence future trends in cholera transmission.  This evolving approach can also be beneficial for policymakers to interpret and respond to potential disease systems. Moreover, our methodology is replicable and adaptable, encouraging future research on disease dynamics.
\end{abstract}





\maketitle

\section{Introduction}\label{sec1}
Cholera persists as a formidable global health challenge, necessitating comprehensive strategies to elucidate, model, and effectively control its transmission dynamics. Cholera is an acute gastrointestinal disease characterized by gram-negative based Vibrio cholerae. Cholera causes extreme watery diarrhea, resulting in fatal dehydration and, consequently, kidney failure, abdominal cramps, vomiting, hypovolemic shock, and death. Transmission of bacterial infection is materialized through the fecal-oral route from contaminated water or food \citep{Nelson20009,pnas05020, Montero2023, LUBY2020A110,pmed}. Vibrio cholerae has emerged as seven important global pandemics during 1817-1824 in India, Asia, and southeastern Africa \citep{Muzembo2022, CAMACHO2018e680, Lopez2020c}. In developing and under-developed countries, poor water supply, lack of sanitation, and bad hygiene practices contribute to its transmission \citep{Mukandavire20}. Rapid outbreaks occurred and highlighted how bacterial pathogen and lytic bacteriophage propelled and quenched the cholera epidemic in Zimbabwe during 2008-2009 \citep{pnas0008}. Recently, Malawi, a landlocked country in southeastern Africa, has experienced the worst cholera outbreak \citep{Migg0o2023}. Nowadays, emerging as well as re-emerging infections like cholera are an open challenge \citep{Morens20004} while environmental reservoir has a significant impact on transmission \citep{Kong20014}. In-apparent infections also apprehend a key to expounding the trend of cholera outbreak \citep{King20008}. In order to mitigate deadly destruction leading to cholera outbreaks \citep{leo2019machine}, diagnosis followed by well-informed decisions as well as interventions are to be taken into account in response to epidemics. Moreover, optimal vaccine allocation can diminish epidemic settings \citep{Moore20150, POSNY201538}. Modeling the dynamics of cholera transmission is a significant endeavor and is still challenging.\par

Recently, various infectious disease modeling approaches have been adopted for modeling the cholera infection pattern \citep{10982021, Acharya2024}. These studies on the cholera epidemic undoubtedly motivate us to recognize the transmission pattern. A recent mechanistic model illustrated prediction ability in Haiti cholera epidemic \citep{do88}. Modeling the aqueous transport of pathogens is also part of cholera transmission \citep{1098304}. Furthermore, transmission via household extremely contributes to the cholera epidemic \citep{Meszaros2020a}. The expected time in anticipating cholera extinction is quantified based on available data in Lusaka, Zambia \citep{Maity20123}. In developing countries, limited resources lead to rapid growth in cholera transmission \citep{Nyabadza20019}. Moreover, public health interventions are of great importance in the mitigation of cholera outbreaks \citep{1098304}. A cost-effective strategy is considerably beneficial to government-undertaken interventions \citep{101371, Das20021, Das20201}. Modeling of optimal intervention strategies noticeably designs a framework for mitigation of cholera outbreaks \citep{Miller2010}. Health organizations can take the initiative to limit the development of serious infectious disease outbreaks in a number of ways by utilizing a forecasting approach \citep{panja2023ensemble, leo2019machine, nino, eeg, chakraborty2019forecasting}.\par

Reliable and accurate forecasting of epidemic data plays a significant role for public health officials in developing effective prevention measures for suppressing epidemic outbreaks like cholera infections \citep{panja2023epicasting}. Various statistical and machine learning models have been designed to provide real-time short-term forecasts of Cholera for Yemen, Haiti, and Bangladesh \citep{nishiura2017transmission, pasetto2018near, martinez2017cholera, daisy2020developing,human_behaviour, GHOSH}. However, these methods do not explicitly learn the mechanistic dynamics and, therefore, cannot give an understanding of how the epidemic will unfold over a longer time horizon. Such long-term trajectory modeling remains the strong suite of mechanistic models that incorporate disease characteristics and an understanding of epidemic progression. Despite their capabilities, compartmental models are not scaled, and calibrating them is prone to noise \citep{hazelbag2020calibration, Kharazmi2021, NING2023106693}. In this interface, there exist hybrid models that combine compartmental models with statistical and machine learning methods \citep{Ye2025, KB2025111988, barman2025epidemic} to generate better short-term and long-term forecasts \citep{ghosh2021integrated, rodriguez2023einns, v15081749}. However, an especially designed epidemic-guided model for a cholera outbreak is missing in the literature.\par

To our knowledge, mathematical modeling and forecasting of cholera epidemics have so far been investigated by incorporating various parameters \citep{ Kong20014, POSNY201538, leo2019machine, Nyabadza20019, de2011forecasting, v15081749, osti1706217}.
Nevertheless, mechanistic model-driven forecasting based on the integration of mechanistic and machine-learning approaches remains undeveloped. For this challenge, we aim to develop the qualitative dynamics of cholera epidemics coupled with real-time cholera cases. Subsequently, we focus on real-time forecasting of cholera outbreaks in order to enhance the remarkable resemblance of transmission dynamics with machine learning models of a cholera epidemic in Malawi as a case study. At the outset, the cholera model is proposed, followed by the derivation of the basic reproduction number ($R_0$). Furthermore, the model is calibrated with the delay rejection adaptive metropolis (DRAM) algorithm using real-time cholera cases in Malawi. In addition, partial rank correlation coefficient (PRCC)-based sensitivity analysis is performed to identify crucial parameters for investigating the dynamics of the cholera epidemic. Asymptotic as well as rich dynamics are explored in the proposed model. In continuation, we develop an epidemic-informed forecasting model by integrating the temporal dynamics of the cholera model (proposed in Section \ref{sec3}) into statistical and machine learning frameworks. Our proposed approaches enable the use of epidemiological information and leverage the superiority of statistical and machine learning models to produce accurate forecasts even in the long term. Through the experimental evaluation of the Malawi cholera dataset, we observe that domain knowledge-based forecasting models lead to more efficient real-time forecasting. These experimental results are further validated using several statistical metrics for its robustness check. \par
 We summarize our contributions as follows:
\begin{itemize}
    \item In this study, we develop a mathematical model, namely SIBR, that consists of susceptible (S), infected (I), vibrio cholerae bacteria (B), and recovered (R) to explore the qualitative dynamics of cholera transmission across human and bacterial populations followed by parametric calibration through real-time cholera cases in Malawi. 
    \item This paper also proposes two epidemic-informed machine learning models (EIML), such as epidemic-informed autoregressive integrated moving average (EI-ARIMA) and epidemic-informed autoregressive neural networks (EI-ARNN). These hybrid models integrate the mechanistic model with machine learning and statistical models for the cholera forecasting task. EIML methods use the knowledge of infection dynamics obtained from the SIBR model in ARIMA and ARNN. This helps EI-ARIMA and EI-ARNN to learn the latent epidemic dynamics and embed that information into the forecasting framework.
    \item Empirical evaluations of the EIML approaches for forecasting the cholera incidence cases of the Malawi region highlight the importance of using epidemic dynamics in the data-driven techniques.
\end{itemize}
The rest of the paper is organized as follows. In Section \ref{sec2}, we develop the mechanistic model and examine its qualitative dynamics, validating our analytical results through numerical analysis. Section \ref{sec3} emphasizes the real-time prediction of cholera cases by developing epidemic-informed machine learning models, conducting experimental evaluations statistically, and performing benchmark comparisons. In conclusion, we present a summary of insights gained from our studies.

\section{Mechanistic model of cholera}\label{sec2}
\subsection{\textbf{Data Source}}
 Data to calibrate the cholera model were collected from the cholera surveillance dashboard under the Ministry of Health, Malawi (https://cholera.health.gov.mw/surveillance). The time span from 28 February 2022 to 10 July 2023 is taken into account. Moreover, weekly cases are considered for a case study.

\subsection{\textbf{SIBR model set-up}}
\begin{figure}[h]
\begin{center}
\includegraphics[width=9cm, height=6cm]{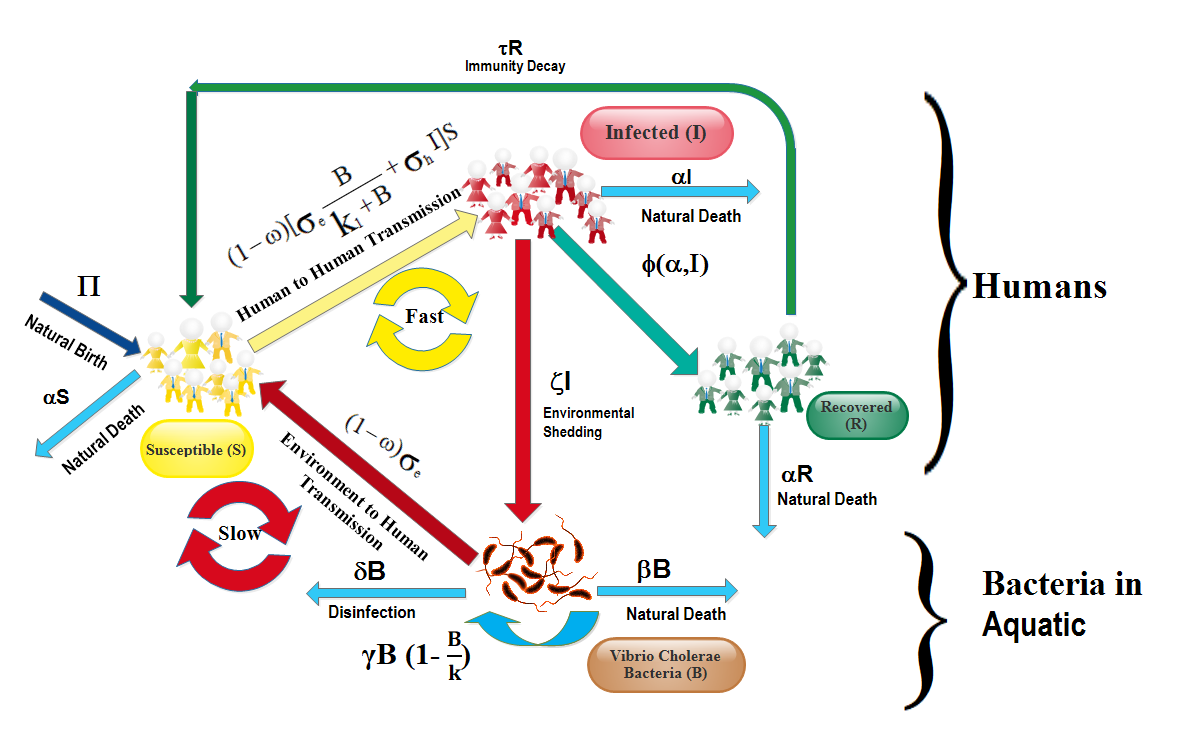}
\caption{~ Schematic portrayal of SIBR model. The flowchart demonstrates the interplay of individuals in the model: susceptible (S), infected (I), vibrio cholera bacteria (B), and recovered (R).}\label{fig1}
\end{center}
\end{figure}

The human population and Vibrio cholerae are believed to represent varied groups within society \citep{Montero2023, Lopez2020c, Montero2023-yc}. Generally, the rate at which cholera is preyed upon can be more accurately depicted in transmission dynamics using a nonlinear response \citep{Mukandavire20, POSNY201538, Maity20123}. Even in situations devoid of predation or human intervention, the intrinsic growth of bacterial populations adheres to a logistic growth model \citep{Kong20014}. In truth, cholera bacteria can be transmitted from environmental sources to humans, as well as between humans themselves \citep{Maity20123,  SUN2017235}. Nevertheless, incorporating either mode of transmission within the human population, alongside the logistic growth of bacterial populations, leaves the dynamics of cholera transmission largely unexamined. Furthermore, the availability of hospital beds influences the recovery rate \citep{POSNY201538, Nyabadza20019}. It is crucial to acknowledge that elements like reinfection due to waning immunity, water sanitation standards, vaccination coverage, and population disinfection must be considered when exploring the complexities of cholera transmission \citep{LUBY2020A110, Meszaros2020a, SUN2017235}. This study excludes cholera-related mortality and immunity acquired through infection to simplify the purpose. Consequently, we propose a system of differential equations to describe the changes in population states over time. The population is categorized into three groups: susceptible individuals ($S(t)$), infected individuals ($I(t)$), and recovered individuals ($R(t)$). The influx of susceptible individuals primarily comes from births and immigration. Cholera infections occur via both infected individuals and polluted water sources. The cholera transmission process is represented through a set of coupled differential equations, as shown in Fig. \ref{fig1}.
    \begin{align} \label{eq1}
\dot{S} = & \; \pi + \tau R- (1-\omega)  \bigg[\frac{\sigma_{e}B}{k_{1}+B} +\sigma_{h}I \bigg]S -\alpha S, \notag\\
\dot{I} = & \; (1-\omega)  \bigg[\frac{\sigma_{e}B}{k_{1}+B} +\sigma_{h}I\bigg]S - \Phi(a,I)I - \alpha I, \notag\\
\dot{B} = & \; \gamma B \bigg(1-\frac{B}{k}\bigg) + \xi I -\beta B -\delta B, \\
\dot{R} = & \; \Phi(a,I)I-(\tau +\alpha) R. \notag
\end{align}
The initial conditions are adopted in model \eqref{eq1} as
\begin{align}\label{eq2}
 &S(t_0)=S_0 \geq 0, I(t_0)=I_0 \geq 0, B(t_0)=B_0 \geq 0,\\
 &R(t_0)=R_0 \geq 0 ~\mbox{with}~ S_0+I_0+R_0 \neq 0.\notag
\end{align}
Table \ref{table1} provides several epidemiological parameters' interpretation and baseline values.

The initial equation in the model \eqref{eq1} illustrates the growth dynamics of susceptible individuals. The average recruitment of new population in terms of births is entered in susceptible individuals at rate $\pi$. Specifically, susceptible individuals can become infected through exposure to contaminated environments and contact with infected individuals at rates $\sigma_{e}$ and $\sigma_{h}$, respectively. Furthermore, individuals who have recovered can be reinfected at a rate of $\tau$. Here, $k_1$ corresponds to the half-saturation density of bacterial predation (< $k$). In the context of bacterial predation, the concentration of bacterial predation follows  Michaelis-Menten kinetics and reaches half of its maximum rate ($k$). The second equation pertains to the dynamics of infected individuals. In this context, susceptible individuals contract the infection and are admitted to hospitals for recovery as defined by the recovery function $\phi$. This function incorporates the number of patients and the bed-to-population ratio $a > 0$ in the hospital to ascertain the recovery rate. The formulation for $\phi(a, I)$ is given by $\phi(a, I) = \phi_{0} + (\phi_{1} - \phi_{0}) \frac{a}{a+I}$, where $\phi_{1}$ represents the minimum recovery rate per capita linked to the availability of sufficient medical resources, the scarcity of infections, and the inherent characteristics of a specific disease; $\phi_{0}$ indicates the minimum per capita recovery rate achievable with minimal healthcare resources \citep{Nyabadza20019}.
The third equation describes how the bacterial population grows over time. Bacteria proliferates logistically at an intrinsic growth rate of $\gamma$ while also considering the environmental carrying capacity $k$, which accounts for the slowing of growth as the population reaches larger sizes due to limited resources. Vibrio cholerae spreads through the environment via infected individuals at a rate $\xi$ during an outbreak. In this context, a disinfection strategy at a rate of $\delta$ is employed as an intervention, such as promoting proper hand hygiene and ensuring access to clean water, which helps disrupt the disease transmission chain. Additionally, $\beta$ denotes the decay rate of the bacteria. The final equation illustrates the growth dynamics of individuals who have recovered.

\subsection{\textbf{Model analysis}}

We foremost study whether the solutions of system \eqref{eq1} are non-negative and bounded. Subsequently, we establish fundamental qualitative properties of the system \eqref{eq1} within $\mathbf{R}^4_+$.
\begin{lema}
With positive initial conditions \eqref{eq2} for the system \eqref{eq1}, the solutions S(t), I(t), B(t), and R(t) remain non-negative for all $t > 0$. 
\end{lema}
The proof of the lemma is given in \ref{lema1}.

\begin{lema}
In the region $\Xi$, the solutions of the SIBR model \eqref{eq1} are bounded uniformly under initial values \eqref{eq2}.
\end{lema}
The proof of the lemma can be obtained in \ref{lema2}.

\subsubsection{The equilibria}\label{equilibria}
Equating zero the right part of system\eqref{eq1}, equilibrium points are derived:
\begin{enumerate}[I.]
    \item Cholera-free equilibrium ($W^{0}$): The cholera-free equilibrium is the state that represents the scenario where there is an absence of cholera in the population. Here, cholera-free equilibrium $W^{0} = (S^0,I^0,B^0,R^0)=\left(\frac{\pi}{\alpha},0,0,0\right)$.
    \item Cholera-present or endemic equilibrium ($W^{*})$: The endemic equilibrium $W^{*}$ = $(S^*,I^*,B^*,R^*)$ at which $I \neq 0$, where 
    \begin{equation*}
        \begin{aligned}
        S^* = & \; \frac{B^*\left(A+\frac{\gamma B^*}{\xi k}\right)\left[\frac{\phi_0+(\phi_1-\phi_0)a} { B^*(A+\frac{\gamma B^*}{\xi k} +\frac{a}{B^*})}+\alpha\right]}{\left(1-\omega\right)\left[\sigma_e \frac{B^*}{k_1+B^*} +\sigma_h B^* \left(A+\frac{\gamma B^*}{\xi k}\right)\right]}, \\
        I^* = & \; B^*\left[A+\frac{\gamma B^*}{\xi k}\right],\\
        R^* = & \; \frac{1}{\left(\tau+\alpha\right)}\left[\phi_0 B^* \left(A+\frac{\gamma B^*}{\xi k}\right) \right. + \\
        & \;\left. \left(\phi_1 - \phi_0\right)a\left(\frac{A+\frac{\gamma B^*}{\xi K}}{A+\frac{\gamma B^*}{\xi k}+\frac{a}{B^*}}\right)\right],
        \end{aligned}
    \end{equation*}
   provided $A = \frac{1}{\xi}\left(\beta+\delta-\gamma \right)>0.$
\end{enumerate}

\subsubsection{Extraction of basic reproduction number ($R_0$)}
In the model, the cholera-free equilibrium $W^0 (S^0, I^0,$ $ B^0, R^0)$ $= \left(\frac{\pi}{\alpha},0,0,0\right)$ represents a condition in which the community or society is devoid of infection. The basic reproductive number $R_0$ gauges the anticipated patterns of outbreaks. Various factors influence $R_0$, including the duration of infectivity in individuals affected, the contagiousness of the pathogen, and the degree of interaction between infected individuals and the susceptible population.\par
Here, $\Tilde{F}$ and $\Tilde{V}$ signify the transmission matrix and the transition or removal matrix, respectively. Specifically, $\Tilde{F}$ accounts for new infections resulting from both direct and indirect pathways that contribute to the spread of the infection. In our analysis, $\gamma B \bigg(1-\frac{B}{k}\bigg)$  plays a role in generating new infections as an environmental reservoir for transmission instead of serving as a means of clearance. Additionally, $\xi I$  signifies new infections arising from infected individuals, serving as an extra resource for infection rather than facilitating the removal or transition of infection. From a theoretical perspective, $\Tilde{F}$ encompasses all potential avenues for new infections to emerge through direct (human-to-human) and indirect (environment-to-human) sources, while $\Tilde{V}$ is solely concerned with the removal or transition processes. To substantiate the correct influence of $R_0$, the terms driven by transmission are positioned within the F matrix. This arrangement guarantees conformity with the next-generation matrix, enabling the derivation of an accurate $R_0$. $\Tilde{F}$ and $\Tilde{V}$ matrix are as follow:

$$ \Tilde{F}=\begin{bmatrix}
(1-\omega)  \bigg[\frac{\sigma_{e}B}{k_{1}+B} +\sigma_{h}I \bigg]S\\
\gamma B \bigg(1-\frac{B}{k}\bigg) + \xi I 
\end{bmatrix},
\quad 
\Tilde{V}=\begin{bmatrix}
 \Phi(a,I)I + \alpha I\\
 \beta B + \delta B
\end{bmatrix}.$$
Here, $R_0$ corresponds to the largest eigenvalue of the next generation matrix $F{V^{-1}}$  where $F=\frac{d\Tilde{F}}{dX}$,  $V=\frac{d\Tilde{V}}{dX}$, $X=[I, B]^{'}$ and $'$ represents transpose of a matrix. So, we have
\begin{equation*}
    \begin{aligned}
R_0 = &\; \frac{1}{2}\bigg[\frac{\sigma_h \pi(1-\omega)}{\alpha(\phi_1 + \alpha)} +\frac{\gamma}{\beta+\delta}+\\
  & \sqrt{\bigg(\frac{\sigma_h \pi(1-\omega)}{\alpha(\phi_1 + \alpha)}+ \frac{\gamma}{\beta+\delta}\bigg)^2+\frac{4\pi(1-\omega)(\xi\sigma_e-k_1\sigma_h\gamma)}{k_1\alpha(\phi_1+\alpha)(\beta+\delta)}}\bigg].
\end{aligned}
\end{equation*}
Here, the feasibility of $R_0$ holds for $(\xi\sigma_e-k_1\sigma_h\gamma) >0$, i.e., basic reproduction number can be obtained for a higher transmission rate of the environment to a human being ($\sigma_e$) than a human being to a human being($\sigma_h$). The additive framework of $R_0$ indicates the existence of various transmission routes. The initial term pertains to the direct (human-to-human) transmission progression from infected individuals, focusing on the interplay of various influencing factors. The subsequent term reflects how the dynamics of progression and clearance in infected individuals impact their recovery or mortality rates, thereby affecting transmission patterns. The radical term incorporates a quadratic correction to the fundamental sum of infection contributors, highlighting interactions that affect transmission. The introduction of an additional term within the square root suggests the role of indirect transmission via an intermediate host. The square root signifies the nonlinear interactions among multiple transmission pathways. This additive framework pinpointed the key contributors to transmission dominance. Nevertheless, the additive composition of $R_0$ is consistent with scenarios in the next-generation matrix, where independent contributions from multiple transmission routes are adjusted through quadratic correction. In next-generation matrix calculations, the joint effects of transmission and progression rates define the basic reproduction number.\par
Here, we perform model calibration followed by numerical results to validate analytical findings through biological interpretations.
\subsection{Parametric calibration}
\begin{figure}[h]
\center
\includegraphics[width=8cm,height=6cm]{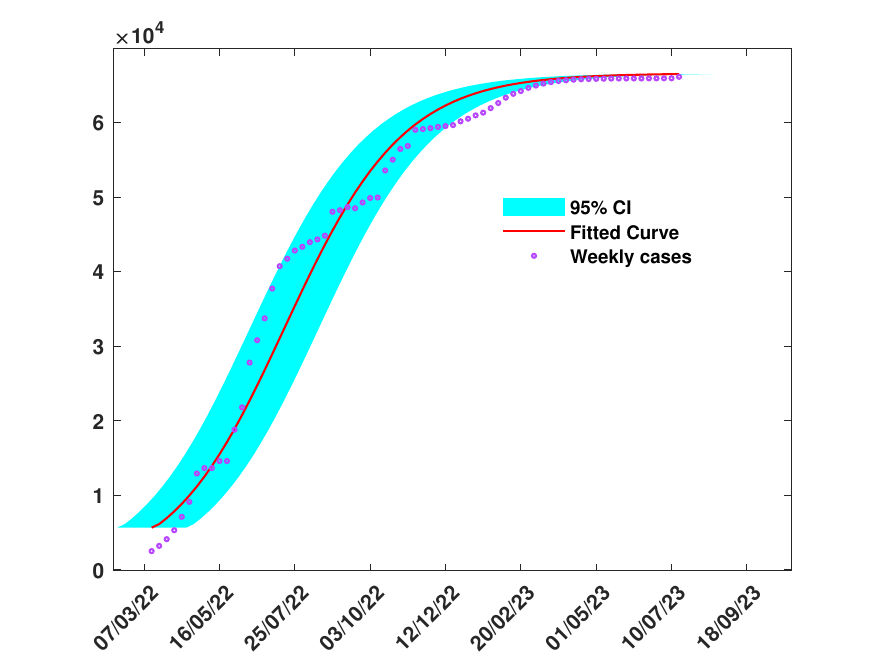}
\caption{~The fitted SIBR model to new cases (weekly) with $95\%$ confidence interval. Cholera data is collected from Country \href{https://cholera.health.gov.mw/surveillance}{Malawi} for the time span from February $22^{nd}$, 2022 to July $10^{th}$, 2023. The 95\% confidence interval for the fitted curve is also plotted here.}
\label{fig2}
\end{figure}

\begin{figure*}[h!]
\center
\includegraphics[width=18cm,height=7cm]{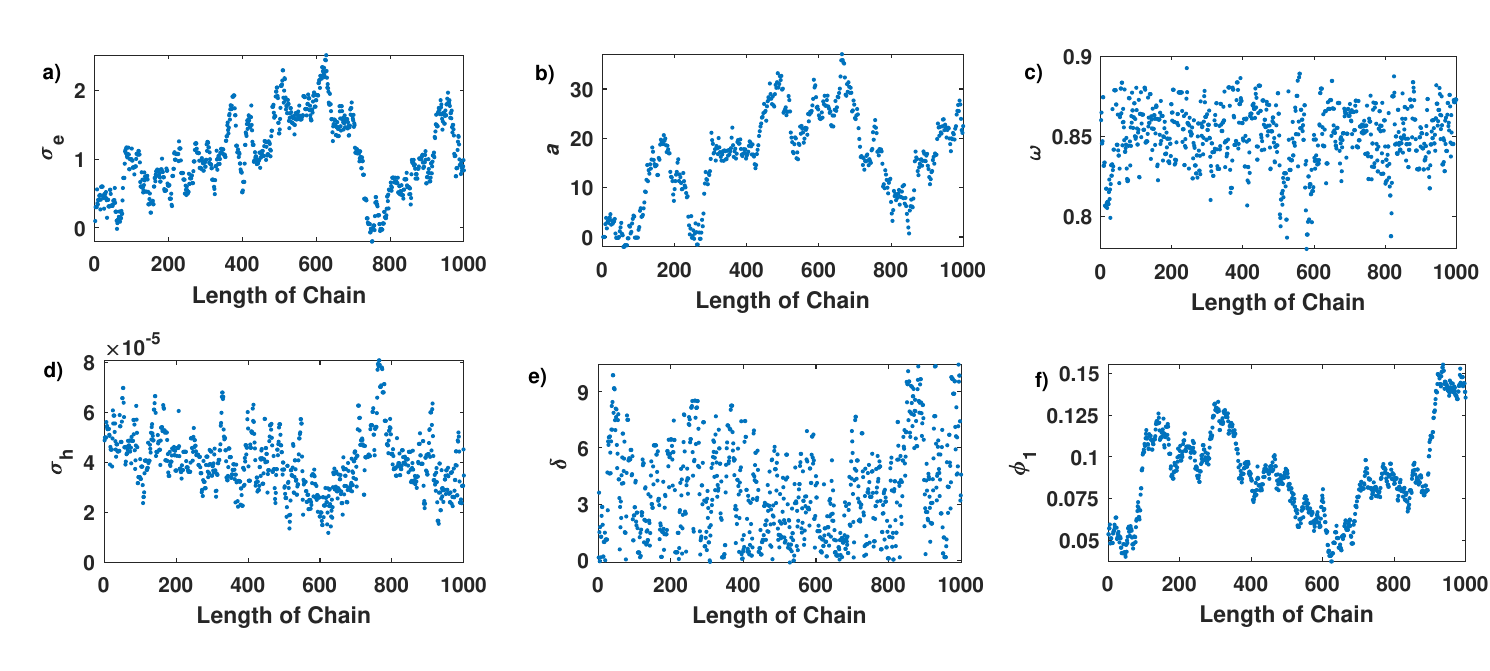}
\caption{~Scatter plot showing 1D DRAM Chain. Here, the mean value is assigned as estimated parameter values from initial mean values, along with the upper and lower limits of the parameters and the length of chain 1000.} Parameters are estimated with an initial guess of $\sigma_e= 0.058$, $\sigma_h= 0.000014$, $\omega=0.86$, $\delta=5$, $\phi_1=0.9$ and $a=10$ with an acceptance rate of $95\%$. Details are given in Table \ref{table3}.
\label{fig3}
\end{figure*}

For calibration of the model, some parameter values from model \eqref{eq1} are estimated and listed in Table \ref{table3}. Additionally, the initial conditions are also estimated and presented in Table \ref{table1}. The nonlinear solver $\it{fminsearch}$ in Matlab is utilized for the model calibration. Furthermore, the Delayed-Rejection Adaptive Metropolis (DRAM) algorithm is used to create a $95\%$ confidence interval. A comprehensive explanation can be found in \citep{haario2006dram}. The fitted curve representing the $95\%$ confidence interval of weekly new cholera cases is illustrated in Fig. \ref{fig2}. The DRAM chains are examined to display the probabilistic values of the model parameters in Fig. \ref{fig3}. In Fig. \ref{fig3}, the DRAM method calculates the mean values for the parameters $\sigma_e$, $\sigma_h$, $\omega$, $\delta$, $\phi_1$, and $a$. The initial mean values, along with the upper and lower limits of the parameters, are shown in Table \ref{table3}.

\begin{table*}[]
\caption{\label{table}The parameters values of the SIBR model \eqref{eq1}.}\label{table1}
\begin{tabular}{llll}
\hline
Parameter & Description &  Value& Reference\\ 
\hline
$\pi=\alpha\times N$& Average recruitment rate&$-$& - \\
$\alpha$& Human natural birth and death rate & 43.5 year$^{-1}$ &\citep{pnas0008}\\
$\tau$&Loss of natural immunity rate&0.5 year$^{-1}$& \citep{King20008}\\
$\omega$&Water sanitation efficacy&-&Estimated\\
$\sigma_{e}$& Transmission rate of environment to human being &day$^{-1}$ &Estimated\\
$\sigma_{h}$& Transmission rate of human being to human being &day$^{-1}$&Estimated\\
$k_{1}$&Half-saturation bacteria predation density ($<k$)&$10^6$ cells/ml&\citep{pmed}\\
$\phi_{0}$& Rate of minimum recovery of human&0.015 day$^{-1}$ &\citep{Nyabadza20019}\\
$\phi_{1}$& Rate of Maximum recovery of humann&day$^{-1}$&Estimated\\
$a$& Ratio of Hospital bed population &day$^{-1}$&Estimated\\
$\gamma$&Pathogen efficiency of maximum per capita Growth &0.2 day$^{-1}$ &\citep{Kong20014}\\
$k$&  Carrying capacity of Pathogen & $10^5$ Cell/Litre&\citep{Kong20014}\\
$\xi$&Shedding rate of Vibrio Cholerae & 10 cells ml$^{-1}$day$^{-1}$ &\citep{pmed}\\
$\beta$&Decay rate of Vibrios &1/30day$^{-1}$&\citep{pmed}\\
$\delta$& Rate of vaccination and dis-infection in population &day$^{-1}$&Estimated\\
\hline
\end{tabular}
\end{table*}

\begin{table}
\caption{Estimation of initial population sizes for model \eqref{eq1}.}\label{table2}
\centering
\begin{tabular}{lll}
\hline
Initial population & Initial value &  Source\\ 
\hline
S(0)&1,97,112 & Estimated \\
I(0)&5,646&Estimated\\
B(0)&21,259&Estimated\\
R(0)&118&Estimated\\
\hline
\end{tabular}
\end{table}
\begin{table}
\caption{Mean values of the estimated parameter values with 95$\%$ confidence intervals for model \eqref{eq1}.}\label{table3}
\centering
\begin{tabular}{lll}
\hline
Parameter & Mean Value &  95$\%$ CI\\ 
\hline
$\sigma_e$&0.0788 & (0.055, 0.082)  \\
$\sigma_h$& 1.36$e^{-5}$&(1.08$e^{-5}$, 1.7$e^{-5}$)\\
$\omega$&0.87&(0.83, 0.89)\\
$\delta$&4&(3.2, 5.5)\\
$\phi_1$&0.091&(0.078, 0.098)\\
$a$&14&(10, 20)\\
\hline
\end{tabular}
\end{table}

A sensitivity analysis, followed by an uncertainty analysis, is performed to determine which parameters are most sensitive to the basic reproduction number and the parameters of the SIBR model.

\subsection{\textbf{PRCC-based sensitivity analysis}}
\begin{figure*}
\center
\includegraphics[width=16cm,height=9cm]{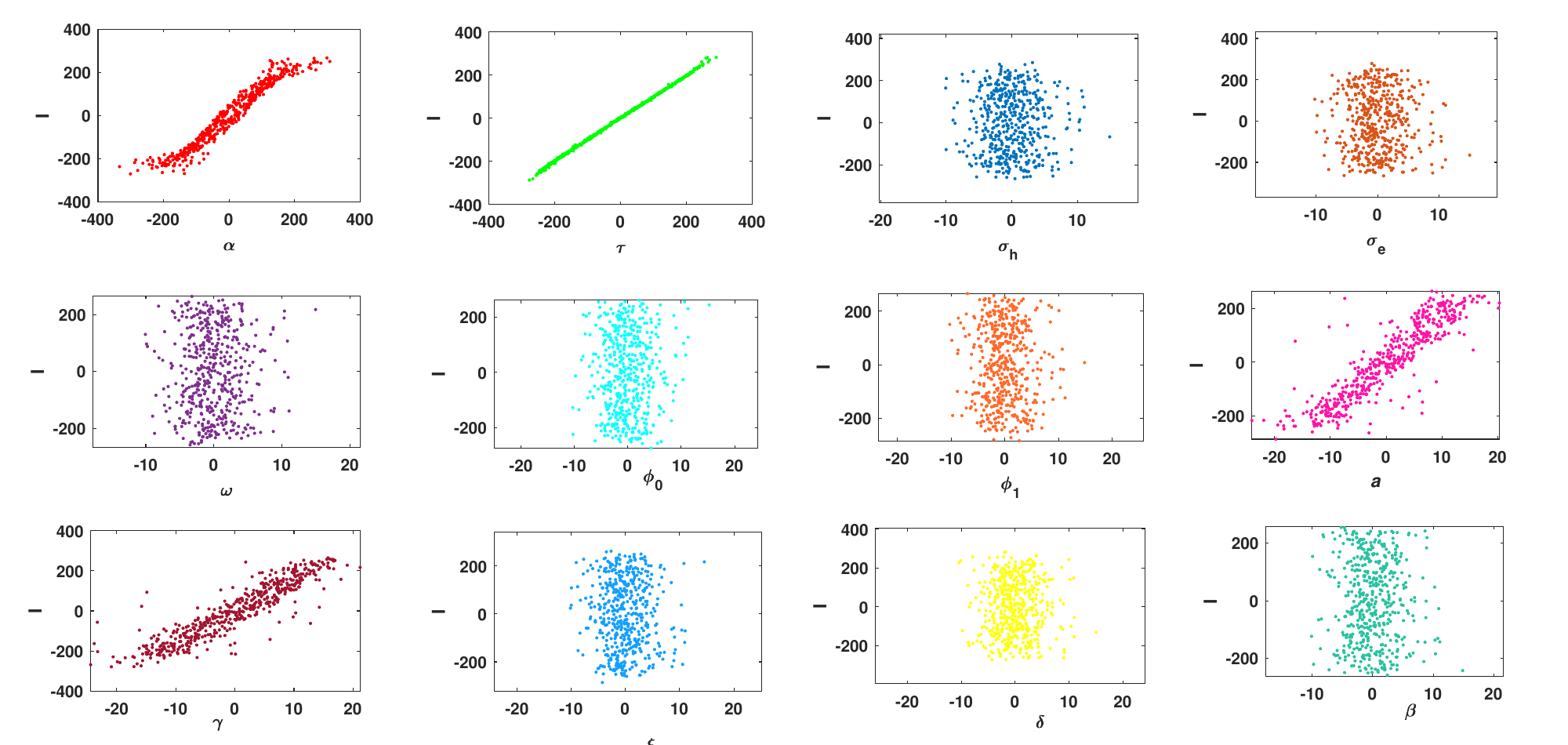}
\caption{~Scatter plots depicting partial rank correlation of parameters to infected individuals ($I$). Here, a randomized sample size of 500 and a unit step size are considered with a significance level of $0.05$. Uniform and N(0,1) probability distributions are employed under the LHS approach. Here the corresponding PRCC value and P-value are $\alpha:(0.96742, 0)$, $\tau: (0.9997, 0)$, $\sigma_h:(0.0313, 0.4853)$, $\sigma_e:(0.0234, 0.6008)$, $\omega:(0.0124, 0.7822)$, $\phi_0:(0.0555, 0.2158)$, $\phi_1:(0.0381, 0.3952)$, $a:(0.8946, 0)$, $\gamma:(0.9123,0)$, $\xi:(0.0523, 0.2376)$, $\sigma_h:(0.0491, 0.2724)$, $\beta:(0.0019, 0.9647)$. Here, $\alpha$, $\tau$, $\gamma$ and $a$ are sensitive parameters for infected individuals ($I$).}
\label{fig4}
\end{figure*}

A sensitivity analysis assesses the statistical effect of uncertainty in system parameters. A helpful method, Latin hypercube sampling (LHS), is utilized to manage parameter uncertainties \citep{marino2008methodology}. The partial rank correlation coefficient (PRCC) offers a transformation of ranks into linear correlation using scatter plots. For the PRCC analysis, seven parameters (specifically, $\alpha, \tau, \sigma_h, \sigma_e, \omega, \phi_0$, and $\phi_1$) are assigned to a standard normal probability distribution, while the other parameters (namely, $a, \gamma, \xi, \delta, \beta$) follow a uniform probability distribution with a sample size of 1000 and a significance level of $0.05$. Additionally, the sign of PRCC indicates a quantitative relationship among the parameters, where a positive sign denotes an increasing correlation and a negative sign represents a decreasing correlation. In Fig. \ref{fig4}, it can be observed that the parameters $\alpha, \tau, a, \gamma$ exhibit a strong correlation with the infected population, meaning these parameters positively influence the spread of infection within the community.\par
\begin{figure}[h!]
\center
\includegraphics[width=8cm,height=6cm]{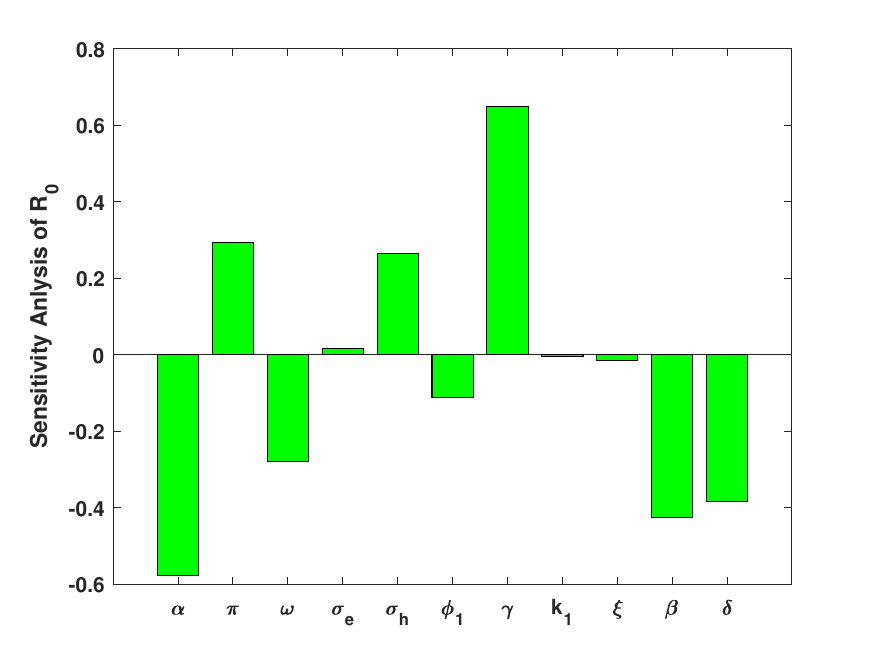}
\caption{~Bar diagram indicating PRCC-based sensitivity indices of parameters to basic reproduction number ($R_0$) with random sample size 500 and significance level 0.05. Here, $\alpha$, $\omega$, $\beta$, and $\delta$ have an inverse relationship with sensitivity, whereas $\pi$, $\phi_h$, and $\gamma$ exhibit a direct relationship with $R_0$.}
\label{fig5}
\end{figure}

In addition, it can also be observed that the sensitivity indices of the basic reproduction number $(R_0)$ in Fig \ref{fig5}. Here, only $\pi, \sigma, \gamma$ are positively as well as remaining negatively correlated to $R_0$.
As the $R_0$ quantifies expected secondary infection, we study the impacts of various parameters on $R_0$, which leads to an increasing or decreasing trend of epidemic evolution.

\subsection{\textbf{Effects of parametric variations on $R_0$}}
\begin{figure*}[h!]
\center 
\includegraphics[width=17cm,height=9cm]{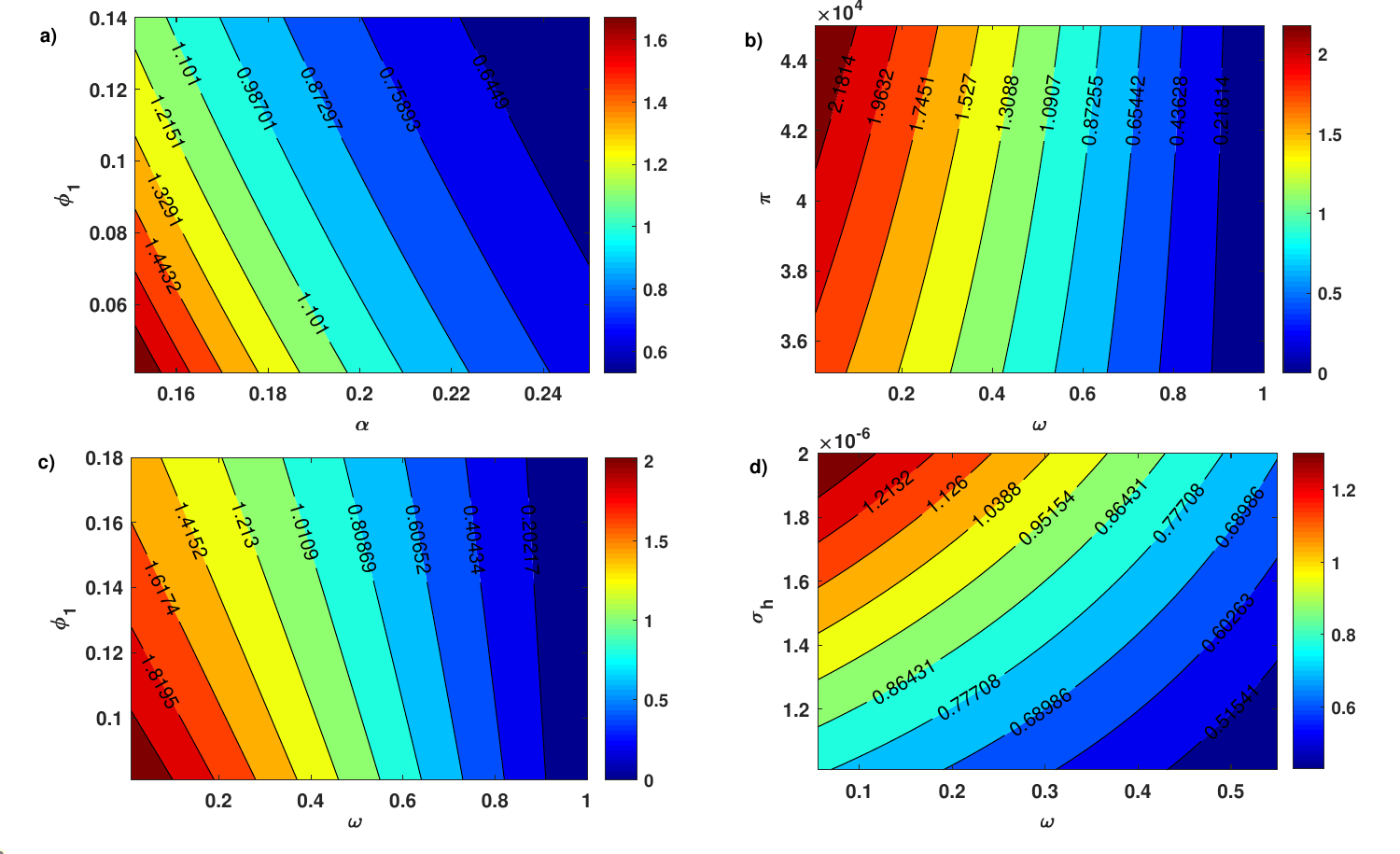}
\caption{~Contour plots showing the changing patterns of basic reproduction number ($R_0$) under parametric planes. Panel a: $R_0$ vs ($\phi_1 \times \alpha$), Panel b: $R_0$ vs ($\pi \times \omega$), Panel c: $R_0$ vs ($\phi_1 \times \omega$) and Panel d: $R_0$ vs ($\sigma_h \times \omega$). In this context, elevated levels of $\pi$, $\sigma_h$, along with reduced values of $\phi_1$, $\alpha$, and $\omega$, influence the dispersal of infection throughout the society..}
\label{fig6}
\end{figure*}

In this section, we analyze the effects on the basic reproduction number, $R_0$, within parametric planes. It is evident that $R_0$ rises as both $\alpha$ and $\omega$ increase within the range of $\phi_1 \times \alpha \in [0.04, 0.14]\times[0.14, 0.26]$ as illustrated in Fig. \ref{fig6}a. This indicates that the rate of secondary infections decreases with a higher maximum recovery rate in hospitals, alongside the birth and death rates of individuals in the population. Additionally, Fig. \ref{fig6}b indicates that $R_0$ growth with an increase in $\pi$, while increases in $\omega$ do not have the same effect within the interval of $\omega \times \pi \in [3.4, 4.6]\times[0.001, 1]$. Here, the occurrence of secondary infections rises with a constant influx of individuals ($\pi$) coupled with a lower water sanitation efficacy ($\omega$). A comparable trend regarding secondary infections is observed in Fig. \ref{fig6}c, which aligns with the findings of Fig. \ref{fig6}a. In Fig. \ref{fig6}c, secondary infections increase as the values of $\phi_1$ and $\omega$ decrease within the parameters $\phi_1 \times \omega \in [0.08, 0.18]\times[0.001, 1]$. In the parametric plane $\sigma_h \times \omega \in [1\times 10^{-6}, 2\times 10^{-6}]\times[0.001, 0.6]$, the rate of secondary infection transmission follows a pattern similar to that shown in Fig. \ref{fig6}d. This suggests that a heightened transmission rate between humans ($\sigma_h$) allows cholera to continue spreading within the community.

Now, we examine qualitative behaviors of the system \eqref{eq1} at biologically feasible equilibrium  $(S^0, I^0, B^0, R^0)$.

\subsection{\bf{Asymptotic dynamics of cholera-free equilibrium}}
\begin{thm}
The SIBR model \eqref{eq1} exhibits local asymptotic stability at disease-free equilibrium $W_0 (S^0, I^0, B^0, R^0)= \left(\frac{\pi}{\alpha},0,0,0\right)$ for $R_0 <1$ as well as unstable $R_0 >1$.
\end{thm}
We omit this as a similar proof is available in \citep{Kong20014, POSNY201538, Nyabadza20019}.

\begin{lema}
   Consider a dynamical system that models the spread of disease: $\frac{dU}{dt}=\Upsilon(U, 0)$, $\frac{dV}{dt}=\Theta(U, V)$, where $\Upsilon(U, 0)$ represents a function of the uninfected subsystem in the absence of infection, and $\Theta(U, V)$ is a function involving both the uninfected ($U$) and infected ($V$) variables. Additionally, a disease-free equilibrium is characterized as $W_0=(U^*, 0)$, with $U^*$ being the equilibrium of the infected subsystem when there is no infection present. If the subsequent conditions are satisfied:
\begin{enumerate}
    \item {\textbf{Stable uninfected subsystem}}: The point $U^*$ is globally asymptotically stable for the equation $\frac{dU}{dt}=\Upsilon(U, 0)$, indicating that the subsystem is stable at $U^*$ when infection is absent.
    \item {\textbf{Degeneration of infected subsystem}}: The function $\Theta(U, V)$ can be expressed in a degenerated form as $\Theta(U, V)=DV-\hat{\Theta}(U,V)$; where $\hat{\Theta}(U,V) \geq 0$ for $(U, V) \in \Re$, and $D=H_{V}\Theta(U^*, 0)$ is regarded as a Metzler matrix (with nonnegative off-diagonal elements) in the region $\Re.$ The matrix $D$ ensures a framework that inhibits oscillatory patterns that might lead to instability at t $W_0$.
    \item {\textbf{Non-negativity of $\Theta(U, V)$}}: In this scenario, $\Theta(U, V) \geq 0$, which indicates that the infected variables tend toward zero over time.
\end{enumerate}
Consequently, the disease-free equilibrium $W_0$ is globally asymptotically stable when $R_0<1$.
\end{lema}

\begin{thm}
The system \eqref{eq1} shows global asymptotic stability at $W^0(S^0,0,0,0)$ if $R_0<1$ in the bounded region $\Xi$.
\end{thm}
 The analytical proof of the theorem is elaborated in Appendix \ref{local stability}. We further examine the qualitative behaviors of the system to illustrate the nonlinear phenomena of cholera transmission dynamics.  
\begin{thm}
The system \eqref{eq1} illustrates asymptotic stability at infection present equilibrium $W^*$ for $R_0>1$. The system further experiences forward bifurcation at $R_0=1.$
\end{thm}
The analytical proof of the theorem is discussed in Appendix \ref{Long-term Dynamics}.

\begin{figure}[h!]
\center
\includegraphics[width=9cm,height=7cm]{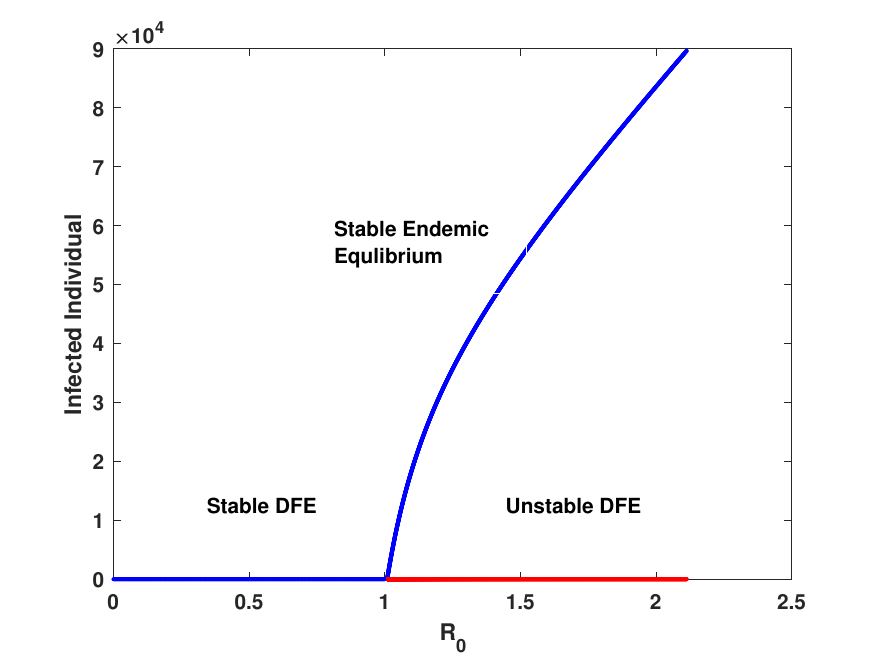}
\caption{~ Basic reproduction number $R_0$ vs infected individual (I) representing forward bifurcation with  $ \pi \in [100, 15000]$. Here, DFE indicates disease-free equilibrium. In the domain of cholera dynamics, the count of newly infected individuals rises with an increase in the recruitment rate.}
\label{fig7}
\end{figure}

The impact of cholera transmission on public health is linked to the basic reproduction number, $R_0$. In this regard, we investigate the local dynamics of the model \eqref{eq1}. In Fig. \ref{fig7}, it is observed that the system undergoes a forward bifurcation. Here, the cholera-free equilibrium ($W^0$) exhibits both stability and instability for $R_0>1$ and $R_0<1$. Additionally, the cholera-present equilibrium is stable when $R_0>1$, with $\pi \in [100, 15000]$ indicating that cholera persists as the population of susceptibles grows in society. In the context of cholera transmission, forward bifurcation highlights the sensitivity of transmission dynamics to changes in parameters. This phenomenon improves policymakers' capacity to predict outbreaks and formulate effective public health strategies to decrease the risk of cholera transmission in vulnerable regions. Recognizing the conditions that instigate forward bifurcation enables health authorities to implement proactive measures to prevent the escalation of disease and successfully oversee public health interventions.
 Our aim now is to investigate the long-term behavior of cholera transmission at the cholera-present equilibrium when $R_0 >1$.
 
\subsection{\textbf{Existence of cholera-present or endemic equilibrium}}
The cholera-present equilibrium ($w^*$) is obtained from the system \eqref{eq1}, as detailed in subsection \ref{equilibria}. To ensure the presence of cholera equilibrium, a sixth-degree equation in $B$ (the bacterial population $B(t)$) is derived as $B^6p_6+B^5p_5+B^4p_4+B^3p_3+B^2p_2+Bp_1+p_0=0$. In this equation, the coefficients $p_i \; (i=0,1,....,6)$ are specified in Appendix \ref{coefficent}. Given the complexity of the equation, we employ Descartes's rule of signs to confirm the existence of at least one positive root. We can ascertain at least one positive root if the leading coefficient or the constant term has a negative sign while the rest of the coefficients are positive. Additionally, it is necessary to have $B(t)>0$ for the cholera-present equilibrium to exist. 

From the third equation in \eqref{eq1}, we identify $I=f(B)=(\beta+\delta-\gamma)B+\frac{\gamma B^2}{k}$. Consequently, we find that $f'(B)=(\beta+\delta-\gamma)+\frac{2\gamma B}{k}>0$, given that $B(t)>0$, indicating that this is an increasing function, with the condition $(\beta+\delta-\gamma)>0$ ensuring the cholera-present equilibrium $W^*$ can exist.

Next, we aim to delve into the topological structure of the system \eqref{eq1} at the cholera-present equilibrium $W^*$.
\subsection{\textbf{Richer dynamics}} 
To explore the intricate dynamics of the system \eqref{eq1}, we calculate the coefficients of the characteristic polynomial from the Jacobian matrix to analyze the occurrence of the Hopf. Let 
$q(\lambda; m)= q_{0}(m)+q_{1}(m)\lambda+q_{2}(m)\lambda^{2}+q_{3}(m)\lambda^{3}+\ldots+q_{n}(m)\lambda^{n}$ be the characteristic polynomial, with $ q_{n}(m)=1$. So we get,
 $$J_{n}(m)=
\begin{bmatrix}
q_{1}(m) &  q_{0}(m)  & \ldots & 0\\
q_{3}(m) &  q_{2}(m) & \ldots & 0\\
\vdots & \vdots & \ddots & \vdots\\
q_{2n-1}(m)  &  q_{2n-2}(m) &\ldots & q_{n}(m)
\end{bmatrix},
$$
where, Det$(J_{j}(m))=A_{j}(m),~j=1,2....n$, $A_{1}(m)=q_{1}(m)$, 
$A_{2}(m)=q_{1}(m)q_{2}(m)-q_{0}(m)q_{3}(m)$, $\makebox[2em]{\dotfill} $.
 
\begin{figure}[h!]
\center
\includegraphics[width=8cm,height=6cm]{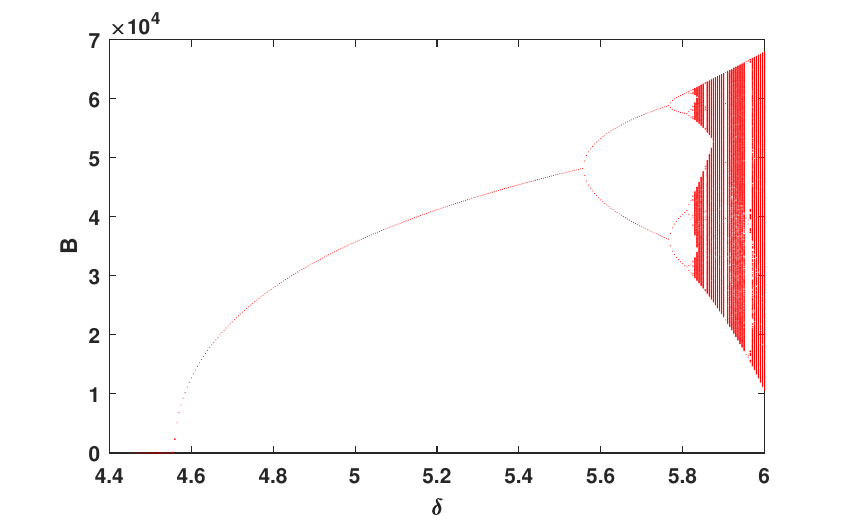}
\caption{~$B_{max}$ vs $\delta$ $\in [4.4, 6]$ showing Hopf as well as period doubling bifurcation. $B_{max}$ is calculated from $B$ solution component of system \eqref{eq1} with time step $\Delta t=0.45$ and initial condition given in Table \ref{table2}. Rich dynamics are depicted, highlighting the non-linear interactions among individuals and new infections that result in the unpredictable growth of cholera cases.}  
\label{fig8}
\end{figure}

The characteristic polynomial of the  system \eqref{eq1} can be obtained as ${\lambda}^{4}+f_{3}(\delta){\lambda}^{3}+f_{2}(\delta){\lambda}^{2}+f_{1}(\delta)\lambda +f_{0}(\delta)= 0$ at $W^{*}$, where 
$f_{3}(\delta), f_{2}(\delta), f_{1}(\delta),$ and $f_{0}(\delta)$ are coefficients. Now, $q_{0}(\delta{*})=[f_{0}(\delta)]_{\delta=\delta^{*}}>0$, $D_{1}(\delta^{*}) =[f_{1}(\delta)]_{\delta=\delta^{*}}$, $D_{2}(\delta{*})=[f_{2}(\delta)f_{1}(\delta)-f_{0}(\delta)f_{3}(\delta)]_{\delta=\delta^{*}}>0 $ and $ D_{3}(\delta{*}) = [f_{1}{\delta}(f_{2}(\delta)f_{3}(\delta)-f_{1}(\delta))-f_{0}(\delta)(f_{3}(\delta))^{2}]_{{\delta}={\delta{*}}}=0 $

and also 
$$
\begin{aligned}
\left.\frac{dD_{3}}{dn_{2}}\right|_{\delta=\delta^{*}}= & -\left[\alpha+\tau-\beta + (1-\omega)  \bigg(\frac{\sigma_{e}B^{*}}{k_{1}+B^{*}} +\sigma_{h}I^{*} \right. +  \\ &  \left.\left(1-\omega\right){\sigma_{h}}S^{*}\bigg)-\left(\phi_{0}+\phi_{1} - \phi_{0}\right) \frac{a^{2}}{(a+I^{*})^{2}} \right]^{2} \\ & \tau (1-\omega)  \left[\frac{\sigma_{e}B^{*}}{k_{1}+B^{*}} +\sigma_{h}I^{*} \right]\\
&\neq 0.
 \end{aligned}
$$
Hence, it can be summarized as a theorem:
  \begin{thm}\label{th7}
The system \eqref{eq1} undergoes a Hopf bifurcation around $W^{*}$ while $\delta$ approaches at the value $\delta=\delta^{*}$ and if following conditions hold: (i) $f_{0}(\delta^{*})>0$ and (ii)$f_{1}{\delta^{*}}(f_{2}(\delta^{*})f_{3}(\delta^{*})-f_{1}(\delta^{*}))=f_{0}(\delta^{*})(f_{3}(\delta^{*}))^{2}.$
\end{thm}  
 In the ongoing analysis, the model \eqref{eq1} experiences Hopf bifurcation, which highlights the long-term dynamics of the cholera population. This suggests that an increase in $\delta$, representing both vaccination and disinfection rates within the community, can lead to unforeseeable non-linear phenomena such as the period-doubling bifurcation illustrated in Fig. \ref{fig8}. The appearance of the period-doubling bifurcation is evident and demonstrates a complex non-linear relationship between individuals and cholera cases. This serves as evidence of the intricate long-term dynamics, indicating unpredictable growth in new infections and reflecting the behavior of cholera transmission. In the realm of cholera transmission, Hopf bifurcation provides important insights into the patterns of outbreaks and the oscillatory behaviors of cholera dynamics. This knowledge is essential for anticipating the erratic nature of outbreaks, which allows for improved preparedness and response to cholera epidemics. By identifying the circumstances that lead to Hopf bifurcations, public health officials can refine their strategies for effectively managing and controlling cholera transmission.\par
We have studied the asymptotic as well as long-time qualitative behavior of the system \eqref{eq1}. To enhance the quality of the investigation, we strengthened the real-time forecasting of cholera cases.

\section{Real-time forecasting of cholera outbreak: A case study in Malawi}\label{sec3}

\subsection{\textbf{Epidemic-informed machine learning model}}
\begin{figure}
    \centering
    \includegraphics[width=8cm,height=4cm]{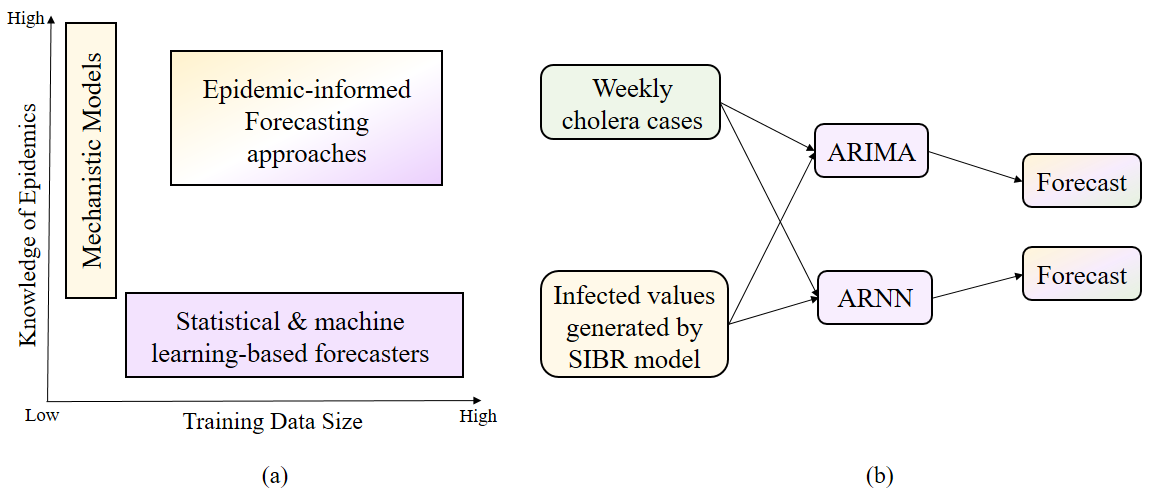}
    \caption{(a) A schematic representation of epidemic-informed machine learning and statistical approach in the context of the epidemic knowledge and the use of the epidemic data size. (b) An illustration of the EI-ARIMA and EI-ARNN models where the infection dynamics from the SIBR model is used as a covariate with the real-world cholera cases in the data science models to generate the forecasts.}
    \label{EI_Model_Fig}
\end{figure}

The epidemic-informed forecasting framework integrates a hybrid approach that combines the compartmental model (specifically the SIBR model discussed in Section \ref{sec2}) with statistical and machine learning techniques. In this research, we mainly focus on two forecasting models that utilize epidemic data for real-time predictions of the cholera outbreak. Our approach merges two distinct modeling methodologies to address their individual shortcomings and leverage their strengths. Specifically, commonly utilized forecasting models have a significant drawback of being referred to as `black-box' data science models (refer to Fig. \ref{EI_Model_Fig}), as they rely solely on historical incidence time series and fail to take into account the scientific processes that influence the disease. In summary, our proposed modeling technique involves the following steps:
\begin{enumerate}[i.]
    \item modeling of a cholera epidemic using SIBR model;
    \item predicting the number of cholera cases using the infection dynamics curve;
    \item stochastic modeling of cholera incidence cases along with the prediction of the mechanistic model as input drivers.
\end{enumerate}
Considering the cholera incidence cases $\left\{Y_t\right\}_{t=1}^N$ from a training dataset of size $N$, indexed by time stamp $t$, along with the estimated infection values ($I_t$) derived from the mechanistic SIBR model, our goal is to predict the future values $\left\{Y_{N+1}, Y_{N+2}, \ldots, Y_{N+h}\right\}$ where $h \geq 1$. We will now detail the construction process of the epidemic-informed autoregressive integrated moving average (EI-ARIMA) and the epidemic-informed autoregressive neural network (EI-ARNN) models. A visual representation of our proposed approach is provided in Figure \ref{EI_Model_Fig}.

\subsubsection{Epidemic-informed autoregressive integrated moving average (EI-ARIMA)}
ARIMA is a widely recognized method for predicting time series data where observations are collected regularly at consistent intervals. This model is made up of three elements: past lagged values $\left(Y_{t-1}, Y_{t-2}, \ldots, Y_{t-p}\right)$, a differencing term to address the non-stationarity in the data, and past lagged errors $\left(\varrho_{t-1}, \varrho_{t-2}, \ldots, \varrho_{t-q}\right)$. The EI-ARIMA($p, d, q, r$) model incorporates the lagged values of $I_t$ as an additional variable in the ARIMA framework and can be mathematically represented as:
    $$Y_t = \sum_{i' = 1}^p  \gamma_{i'} Y_{t-i'} + \sum_{j' = 1}^r  \beta_{j'} I_{t-j'} + \sum_{k' = 1}^q \theta_{k'}\varrho_{t-k'} + \epsilon_t,$$
where $\epsilon_t$ is an iid white noise and $\{\gamma_{i'}, \beta_{j'}, \theta_{k'}\}$ are regression coefficients and have their usual interpretations. To handle non-stationarity, differencing of order $d$ is applied to $Y_t$ before fitting the model \citep{hyndman2018forecasting}. 

\subsubsection{Epidemic-informed autoregressive neural network (EI-ARNN)}
The EI-ARNN($u, v, r$) model treats the future values of the cholera time series as a non-linear function derived from its past lagged observations and the $I_t$ values obtained through the SIBR model. This can be expressed mathematically as  
\begin{equation}\label{EI_ARNN_Eqn}
    Y_t = f\left(Y_{t-1}, Y_{t-2}, \ldots, Y_{t-u}, I_{t-1}, I_{t-2}, \ldots, I_{t-r}\right) + \epsilon_t, 
\end{equation}  
where $\epsilon_t$ represents an independent identically distributed white noise, and $f$ denotes an autoregressive neural network (ARNN) \citep{faraway1998time} featuring $v$ hidden neurons within a single hidden layer. This model integrates both historical data and epidemiological information as its input variables within the neural network architecture. By expanding Eq. (\ref{EI_ARNN_Eqn}) and including the neural network weights $\alpha, \beta$, along with a sigmoidal activation function $g$($\cdot$) used in the ARNN, we can express $f$($\cdot$) as
\begin{align*}
  f&\left(Y_{t-1}, Y_{t-2}, \ldots, Y_{t-u}, I_{t-1}, I_{t-2}, \ldots, I_{t-r}\right)\\
    &= \beta_0 + \sum_{j = 1}^v \beta_j g\left(\alpha_{0, j} + \sum_{i = 1}^u \beta_{i, j}Y_{t-i} + \sum_{i = 1}^r \alpha_{i, j}I_{t-i}\right).
\end{align*}
To ensure a stable learning mechanism in the proposed EI-ARNN framework, we set the number of hidden neurons ($v$) as $v = \lceil \frac{u+r+1}{2} \rceil$ and select the number of lagged inputs $u, r$ by minimizing the Akaike information criterion (AIC) \citep{panja2023ensemble}.

\subsection{\textbf{Experimental evaluation}}
In this part, we assess the effectiveness of the proposed EI-ARIMA and EI-ARNN methods in predicting cholera cases in the Malawi region. The next section offers an in-depth analysis of the statistical and global characteristics of the cholera incidence datasets (see Section \ref{Global_Feat}), a summary of the benchmark forecasting methods from various frameworks employed in the empirical study (refer to Section \ref{base_Fore}), and the various performance metrics used to evaluate the forecasters (see Section \ref{Perf_Measure}). In Section \ref{Bench_Comp}, we present a thorough discussion of the forecasting accuracy of the proposed models in comparison to the leading frameworks.

\subsubsection{Global features of the cholera dataset}\label{Global_Feat}
The total number of weekly cholera incidence cases in the Malawi region, sourced from \url{https://cholera.health.gov.mw/surveillance}, has no missing data and consists of seventy-two observations collected between February 28, 2022, and July 10, 2023. For the purpose of experimental evaluation, we utilize the initial sixty observations gathered from February 28, 2022, to April 17, 2023, to train the forecasting models and evaluate their performance on the multi-step forecast for the subsequent twelve weeks, covering the period from April 24, 2023, to July 10, 2023. The training dataset used in our analysis has an average weekly incidence value of 44425, with a minimum case count of 2007 and a maximum case count of 65833. The coefficient of variation, which measures the relative dispersion of the observations around the mean, is $47.53\%$, indicating a significant level of variability in the cholera incidence rate over the years. Furthermore, to uncover the structural patterns within real-world cholera incidence cases, we examine the following global characteristics of the dataset:

\begin{itemize}
    \item \textit{Stationarity} is useful to test the level or trend stationarity of the cholera incidence time series, Kwiatkowski-Phillips-Schmidt-Shin (KPSS) test is performed using the \textit{kpss.test} function of the `tseries' package inbuilt in \textbf{R} statistical software.
    \item \textit{Seasonality} can identify the seasonal behavior in cholera incidence cases, we conduct the Kruskal-Wallis test using the \textit{kw} function from the 'seastests' package in \textbf{R}.
    \item \textit{Long range dependency} determines the self-similarity or long-range dependency in the cholera cases, we compute the Hurst exponent employing the \textit{hurstexp} function of the `pracma' package in \textbf{R}.
\end{itemize}

\begin{figure*}[h!]
\center
\includegraphics[width=5.4cm,height=4cm]{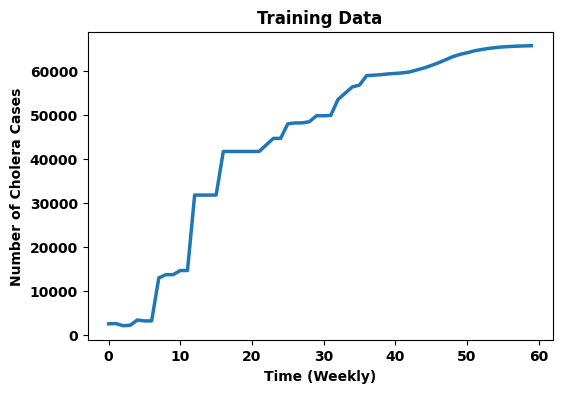}
\includegraphics[width=5.4cm,height=4cm]{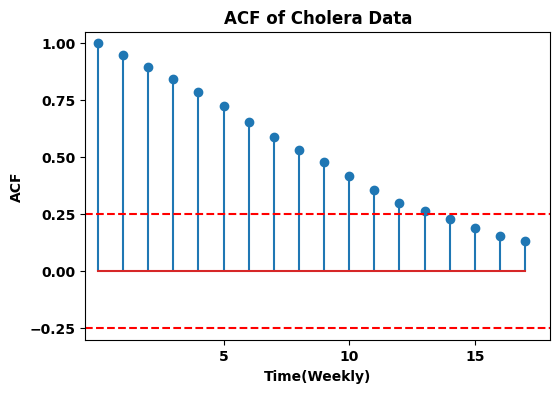}
\includegraphics[width=5.4cm,height=4cm]{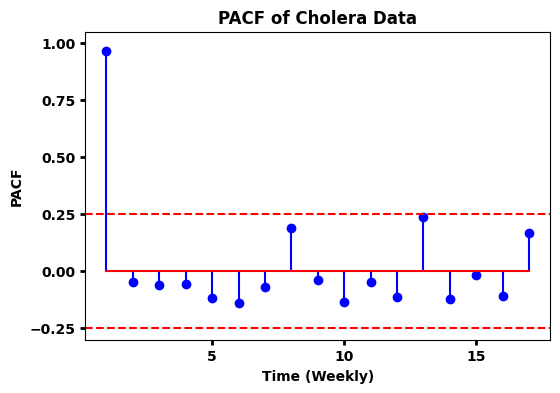}
\caption{~ Illustration of different Plots: training data, autocorrelation function (ACF), and partial autocorrelation function (PACF). The PACF plot shows significant spikes at lag 1, indicating that this lag is important in modeling the underlying data. The ACF plot suggests a non-stationary series with no periodic fluctuations. The overall characteristics of the dataset are long-term dependent, Non-stationary, Positive Trend (0.99), and Non-seasonal.}
    \label{tab:dat_Char}
\end{figure*}

Additionally, we provide the curve of real-time training data, along with the autocorrelation function (ACF) and partial autocorrelation function (PACF) plots for the cholera incidence cases that exhibit global characteristics, as illustrated in Fig. \ref{tab:dat_Char}. Statistical analyses are performed to assess whether the features of cholera cases reveal non-stationarity, linear relationships, long-range dependencies, and non-seasonal patterns in the time series of incidence. Furthermore, the ACF plot in Fig. \ref{tab:dat_Char} suggests that the autocorrelation among the lagged observations of cholera incidence gradually diminishes, becoming statistically insignificant after twelve lagged values. Conversely, the PACF plot indicates a significant correlation at $lag1$ for cholera cases, followed by correlations that lack statistical significance, as shown in Fig. \ref{tab:dat_Char}.

\subsubsection{Baseline forecasters}\label{base_Fore}
In the experimental analysis, we compare the performance of the EI-ARIMA and EI-ARNN approaches with several baseline forecasters, like  RWD \citep{entorf1997random}, ARIMA \citep{box2013box}, ETS \citep{hyndman2008forecasting}, Theta \citep{assimakopoulos2000theta}, TBATS \citep{de2011forecasting}, and ARNN \citep{faraway1998time}. To implement these,  we use the specific functions of the `forecast' package built in \textbf{R}. The AR model is trained using the \textit{ar} function from the `stats' package, and for the Setar model, we employ the \textit{setar} function from the `tsDyn' package through the same statistical software.

\subsubsection{Performance measures}\label{Perf_Measure}
To compare the performance of the different forecasting methods, we consider symmetric mean absolute percentage error (SMAPE), mean absolute percentage error (MAPE), mean absolute scaled error (MASE), mean absolute error (MAE), and root mean squared error (RMSE) metrics. Mathematically, these measures can be expressed as:
{\footnotesize
\begin{equation*}
    \begin{aligned}
        \mbox{ MAPE} & =  \frac{1}{h}\sum_{i =1}^{h} \frac{|Y_{N+i} - \hat{Y}_{N+i}|}{Y_{N+i}}, \quad 
        \mbox{ SMAPE} = \frac{1}{h} \sum_{i=1}^h \frac{2|\hat{Y}_{N+i} - Y_{N+i}|}{|\hat{Y}_{N+i}|+ |Y_{N+i}|} \\
        \mbox{ MAE}  & = \frac{1}{h}\sum_{i =1}^{h} |Y_{N+i} - \hat{Y}_{N+i}|,  \quad 
        \mbox{ MASE} = \frac{\sum_{i = 1}^{h} |\hat{Y}_{N+i} - Y_{N+i}|}{\frac{h}{N-1} \sum_{i = 2}^N |Y_i - Y_{i-1}|}, \\
        \mbox{ RMSE} & = \sqrt{\frac{1}{h}\sum_{i = 1}^{h} (Y_{N+i} - \hat{Y}_{N+i})^2}, 
    \end{aligned}
\end{equation*}}
\noindent where $N$ denotes the size of the training data, $h$ is the forecast horizon, $\hat{Y_t}$  represents the forecast compared to the actual value $Y_t$. According to convention, the minimum measuring value of these performances indicates the 'best' model.
\begin{table*}[]
    \centering
\caption{~Multi-step ahead forecast performance comparison of the baseline models with the proposed EI-ARIMA and EI-ARNN approaches based on different key performance indicators. The least values of the metric (best performance) are \textbf{highlighted}. The results for EI-ARIMA are underlined in the table.}
    \label{tab:Forecast_Result}
    \begin{tabular}{|c|ccccc|} 
\hline Model & MAPE & SMAPE & MAE & MASE & RMSE \\ \hline
RWD & 10.4854 & 9.8292 & 6911.9181 & 628.3562 & 7831.8450 \\
AR & 9.2991 & 9.8611 & 6129.7125 & 557.2466 & 6767.4248 \\
ETS & 4.4267 & 4.3088 & 2918.0093 & 265.2736 & 3250.0273 \\
Theta & 5.4543 & 5.2704 & 3595.4619 & 326.8602 & 4076.6988 \\
SETAR & 1.8565 & 1.8767 & 1223.7181 & 111.2471 & 1318.8838 \\
TBATS & 4.7759 & 4.6399 & 3148.1547 & 286.1959 & 3492.7880 \\
ARIMA & 10.4854 & 9.8292 & 6911.9181 & 628.3561 & 7831.8450 \\
EI-ARIMA (Proposed) & \underline{4.1085} & \underline{4.2013} & \underline{2708.005} & \underline{246.1820} & \underline{2806.8010} \\
ARNN & 1.0985 & 1.1055 & 724.0652 & 65.8241 & 779.0374 \\
EI-ARNN (Proposed) & \textbf{0.3560} & \textbf{0.3567} & \textbf{234.6849} & \textbf{21.3349} & \textbf{237.2166} \\
\hline
\end{tabular}
  \end{table*}

\subsubsection{Benchmark comparison}\label{Bench_Comp}
\begin{figure*}[h!]
    \centering
    \includegraphics[width=14cm, height=6cm]{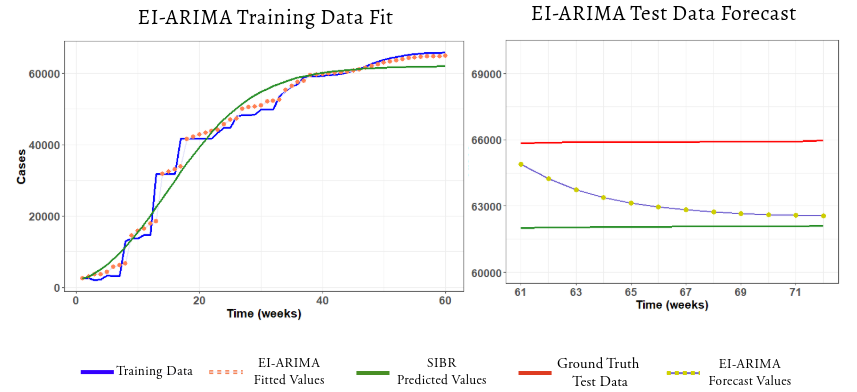}
    \caption{~ Illustrating the time evolution of cholera cases, predicted curve of SIBR, and predicted values of the proposed EI-ARIMA model for both training (60 timestamps) and test data (12 timestamps). Here, \textit{auto.arima} function of the `forecast' package in software, \textbf{R} is applied in EI-ARIMA(1,0,0,1) with \textit{lag1}. Forecasts generated by EI-ARIMA and the actual test data (right) showcase the superior performance of the proposal as compared to the mechanistic model.}
    \label{fig:EI_ARIMA}
\end{figure*}

In this section, we examine the application of epidemic-informed statistical and machine learning methods and emphasize the effectiveness of our proposal against established benchmarks. The EI-ARIMA and EI-ARNN methods primarily integrate the epidemiological insights from the compartmental SIBR model with data-driven techniques to produce precise forecasts of cholera incidence in the Malawi region. Following the estimation of infected cases ($I_t$) derived from the mechanistic SIBR model, we utilize this information as input drivers for statistical and machine learning forecasting techniques. We rebuild the SIBR model with only training data and generate the predicted curve (as seen in Fig. \ref{fig:EI_ARIMA}) for both training and test data. This helps prevent data leakage and overfitting problems in hybrid frameworks. For the EI-ARIMA model, we employ the \textit{auto.arima} function from the `forecast' package in \textbf{R}, incorporating $I_t$ as an additional input. This approach utilizes automated parameter optimization, resulting in the EI-ARIMA(1,0,0,1) model with one lagged input from both the $Y_t$ and $I_t$ series. In the case of the EI-ARNN approach, we implement a stable neural network architecture using the \textit{nnetar} function of the `forecast' package in \textbf{R}. The EI-ARNN model begins with an input layer comprising $u + r$ nodes, followed by a single hidden layer containing $v$ nodes and concluding with an output layer. The input layer processes $u$ lagged values of $Y_t$ and $r$ lagged observations of $I_t$ to create a one-step-ahead forecast. The multi-step-ahead predictions from the EI-ARNN model are generated iteratively. To facilitate a stable learning process, the values of $u$ and $r$ are determined by minimizing the AIC, with $v = \lceil \frac{u+r+1}{2} \rceil$ established as a consideration. In our experiments, we utilize the EI-ARNN(1, 2, 1) model, which includes one lagged value for each of $Y_t$ and $I_t$, along with two hidden nodes to forecast cholera incidence cases in the Malawi province.\par
\begin{figure*}[h!]
    \centering
    \includegraphics[width=14cm, height=6cm]{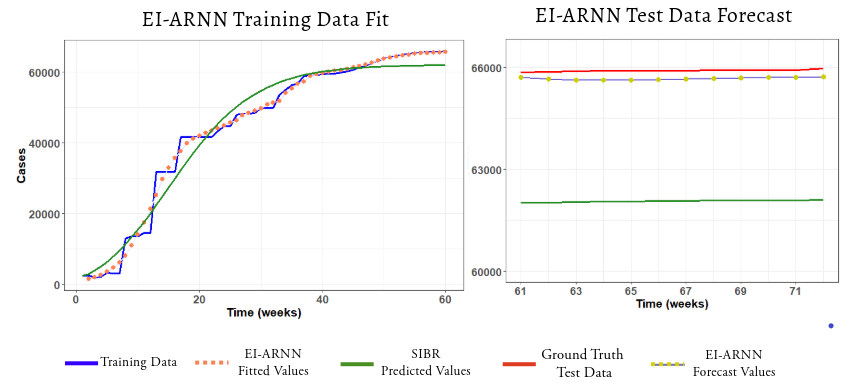}
    \caption{~Illustrating the time evolution of cholera cases, predicted curve of SIBR, and predicted values of the proposed EI-ARNN model for both training (60 timestamps) and test data (12 timestamps). Here, \textit{nnetar} function of the `forecast' package in software, \textbf{R} is applied in EI-ARNN(1, 2, 1) model with \textit{lag1}. Forecasts generated by EI-ARNN and the actual test data (right) showcase the superior (`best')  performance of the proposal as compared to all the benchmark forecasting models considered in this study.}
    \label{fig:EI_ARNN}
\end{figure*}

We apply advanced statistical and machine learning forecasting models to the cholera datasets. While these data-centric approaches effectively capture trends by utilizing historical incidence data, they do not account for important epidemiological factors. The experimental results presented in Table \ref{tab:Forecast_Result} demonstrate that our proposed EI-ARIMA and EI-ARNN architectures outperform the data-centric benchmarks. These improvements are primarily due to the incorporation of prior epidemic knowledge from the mechanistic SIBR models as auxiliary information. Additionally, the stable learning architecture of EI-ARIMA and EI-ARNN helps mitigate overfitting, enhancing the models' generalizability for both short-term and long-term forecasting tasks. Figures \ref{fig:EI_ARIMA} and \ref{fig:EI_ARNN} showcase the cholera forecasts for the Malawi region generated by the EI-ARIMA and EI-ARNN models, respectively. These forecasts, which integrate epidemic knowledge from the SIBR model with historical cholera incidence data, effectively capture the dynamics of cholera transmission. As illustrated in Table \ref{tab:Forecast_Result}, Figure \ref{fig:EI_ARIMA}, and Figure \ref{fig:EI_ARNN}, our forecasting models accurately reflect cholera transmission trends over short timeframes. These forecasts are valuable for public health officials to make timely decisions and can be tracked to assess the effectiveness of intervention strategies. Additionally, the reduced training time required by these models allows for real-time updates based on the most recent data. Hence, they provide an efficient framework for real-time epidemic forecasting, thus allowing policy adjustments during an ongoing outbreak. Overall, the performance of the proposed forecasting methods outperforms the benchmarks and provides accurate cholera forecasts for a short-term horizon. However, these models might struggle to handle sudden peaks in epidemic data. Therefore, further modifications are necessary to improve their performance in such scenarios.


\section{ Concluding remarks}\label{sec5}
The cholera epidemic, recognized as the seventh pandemic, has impacted all aspects of human existence and highlights our susceptibility to significant health risks. A lack of understanding regarding effective prevention and treatment methods results in higher mortality rates from epidemics in developing nations. Given the reported cases during the cholera outbreak, accurate forecasting becomes a vital element for managing transmission patterns through healthcare efforts that operate under resource constraints. Analyzing transmission dynamics and making short-term predictions present a promising but highly challenging approach.\par
In this study, we examined the transmission dynamics of the cholera epidemic and subsequent forecasting using epidemic-informed machine learning models. For practical application, we chose Malawi as a region for the parametric calibration of the cholera model. In this context, we investigated the proposed cholera model by analyzing its asymptotic behavior regarding transmission dynamics within a mathematical framework. The calibration of the model was carried out using the Monte Carlo Markov Chain algorithm. Additionally, a PRCC-based sensitivity analysis was conducted to identify key parameters. The influence of parametric planes on the basic reproduction number ($R_0$) was also demonstrated. We observed forward bifurcation, which implies stable disease transmission when $R_0>1$. Furthermore, the SIBR model underwent period-doubling bifurcation, reflecting complex dynamics regarding Vibrio cholerae. To enhance the investigation, we developed EIML models for real-time cholera epidemic forecasting. Achieving reliable predictions of epidemic trends may prove to be more crucial than just monitoring reported cases. We addressed challenges by integrating ARNN and ARIMA models with mechanistic dynamics through the SIBR model. 
We highlighted the associated quantitative performances to demonstrate the importance of the proposed approach. According to numerical results, EI-ARNN achieves a higher level of accuracy for forecasting transmission trends over a short duration. This comprehensive research could play a vital role in paving a new path for ML-driven epidemic forecasting. This study may be beneficial for exploring the potential of integrating epidemic dynamics with ML models for social good. We can conclude that the EI-ARNN approach is valuable in assisting government officials in making informed decisions regarding public health interventions.\par

A thorough examination of combined epidemic and machine learning models has been presented for making real-time forecasts. Experimental trials have also been carried out. The proposed hybrid models have demonstrated improved accuracy and are recommended for minimizing cholera cases based on insights and trends derived from real-time data. These predictive models help identify transmission patterns, enabling public interventions and effective preventive measures related to water sanitation and disinfection for the community. The EIML models emphasize the development of user-defined functions that offer practical recommendations for lowering cholera incidence. A crucial recommendation for establishing a cholera-free environment with zero fatalities is to maintain distance from household contacts to effectively implement isolation practices. Moreover, enhancing the availability of hospital beds in the community, along with the effectiveness of water sanitation measures, can contribute to the overarching goal of achieving cholera-free regions.\par

In this prospective investigation, our goal was to develop forecasting methods by integrating temporal dynamics into machine learning models. This initiative aimed to facilitate a more in-depth analysis by employing real-time cases along with mechanistic model-driven time series. This thorough approach proves effective in creating sophisticated hybrid ML models. Additionally, subsequent research could explore the application of state-of-the-art datasets to enhance mechanistic models, ultimately leading to a better comprehension of the complexities associated with cholera epidemics. However, this study presents certain constraints, such as the limited number of cholera cases. This constraint may hinder the model's capacity to accurately capture cholera incidents. To address these limitations, future studies will focus on more comprehensive datasets that demonstrate a stronger correlation with cholera transmission. Acknowledging these restrictions, the study highlights the opportunity for sustainable data improvement and has developed a predictive ML model that can support cholera prevention and control measures. Furthermore, RNN and LSTM algorithms could be utilized to construct hybrid EI-ML models to predict future trends in infectious diseases. Apart from modeling the cholera disease trajectory, as investigated in this study, another potential area of research would be to explore how the proposed hybrid frameworks can be adopted for modeling epidemics with different dynamics.

\appendix
\section*{Appendix}

\section{Proof of the Lemma 1}\label{lema1}
\begin{proof}
In order to proof the non-negativity of the system \eqref{eq1}, consider the proposed system \eqref{eq1} as vector form,
$\dot{V} =\psi (V(t))$,
where $V(t)=(v_1, v_2, v_3, v_4)'= (S(t) ,I(t), B(t), $ $ R(t))'$, $V(0)=(S(0), I(0),$ $ B(0), R(0))'\in \mathbf{R}^4_+$
and 
$\psi(V(t))=
\begin{bmatrix}
\psi_{1}(V(t))\\
\psi_{2}(V(t))\\
\psi_{3}(V(t))\\
\psi_{4}(V(t))
\end{bmatrix}
\quad
=\begin{bmatrix}
 \pi + \tau R- (1-\omega)  \bigg[\frac{\sigma_{e}B}{k_{1}+B} +\sigma_{h}I \bigg]S -\alpha S\\
(1-\omega)  \bigg[\frac{\sigma_{e}B}{k_{1}+B} +\sigma_{h}I\bigg]S - \Phi(a,I)I - \alpha I\\
\gamma B \bigg(1-\frac{B}{k}\bigg) + \xi I -\beta B -\delta B \\
 \Phi(a,I)I-(\tau +\alpha) R
\end{bmatrix}.$
Here, $\psi_{i}(V(t))|_{v_{i}=0} \geq 0$ (for $i = 1, 2,..,4$) with the initial condition $V(0) \in \mathbf{R}^4_+ $. By Nagumo's theorem \citep{z90874}, any solutions, say$ V(t) = V(t,V_{0})$, such that $V(t) \in \mathbf{R}^4_+ $ for all $t > 0$ of \eqref{eq1} with initial point $V(0) = V_{0} \in \mathbf{R}^4_+ $ remain positive throughout the region $ \mathbf{R}^4_+.$
\end{proof}
\section{Proof of the Lemma 2}\label{lema2}
\begin{proof}
To establish boundedness, we derive a bounded region within the framework of population as well as bacterial dynamics. In population dynamics, we differentiate $N=S+I+R$ with respect, $t$ which gives
    $\frac{dN}{dt}=\pi-\alpha N$, and follows that $\lim\limits_{t \rightarrow \infty }\sup N(t) =\frac{\pi}{\alpha}$.
We consider the third equation of model \eqref{eq1} for extracting bounded regions in bacterial dynamics. We take,
\begin{align*}
\frac{dB}{dt} &=  \gamma B \bigg(1-\frac{B}{k}\bigg) + \xi I -\beta B -\delta B \\
&=-\frac{\gamma B^2}{k} + (\gamma-\beta-\delta)B +\xi I \quad\mbox{(By simplifying)}\\
& \leq -\frac{\gamma B^2}{k} + (\gamma-\beta-\delta)B +\xi.1 \quad\mbox{( As $I \geq 0$).}
\end{align*}
 Let $T(B)$ be function of bacterial population, whereas $T(B) = -\frac{\gamma B^2}{k} + (\gamma-\beta-\delta)B +\xi$.   
$T(0)=\xi ~\mbox{at}~ B=0.$
Additionally, T(B)=0 has two roots in the form  of
$$B_{1,2} = \frac{k}{2\gamma}\bigg(\gamma-\beta-\delta)\mp\sqrt{(\gamma-\beta-\delta)^2+4\frac{\gamma}{k}\xi}\bigg).$$
As $(\gamma-\beta-\delta)^2+4\frac{\gamma}{k}\xi>0$, it can be obtained $B_2>0$, whereas $B_1<0$. Moreover,
$B(t)\geq 0, ~\forall ~t>0$. It can be concluded that $0 \leq B(t) \leq B_2, ~\forall~ t>$0. Hence, bacterial population size $B(t)$ is bounded above. Thus, the solutions of system \eqref{eq1} remain nonnegative and bounded within the region $\Xi$. Here, $\Xi$ can be defined as 
$\Xi= \{ (S, I, B, R) \in \mathbb{R}^{4}_{+}, 0 \leq B(t) \leq B_2 ~ \& ~  0 \leq S, I, R \leq \frac{\pi}{\alpha}\}.$ This proof establishes the biologically feasible region $\Xi$, where the level of $B(t)$ maintain to provide its roots of $T(B)$ with $B_2$ as optimum bacterial population. This assures the relevance of the model in population dynamics for restricting unrealistic growth of population size.

\end{proof}

\section{Proof of the Theorem 2}\label{local stability}

\begin{proof} 
The model \eqref{eq1} can be represented as
    $\frac{dU}{dt}=\Upsilon(U, V), $ $
    \frac{dV}{dt}=\Theta(U, V),$
where $\Theta(U,0)=0$, $U=(S,R)\in \mathbb{R}^2$ is represented by susceptible or uninfected individual and $V=(I,B) \in \mathbb{R}^2$ is represented by infected individual. Here, $W_0(S^0,0,0,0)$ is disease-free equilibrium of the system \eqref{eq1}. Now,
\small{
$$
\Upsilon(U,V)=\begin{bmatrix}
\pi + \tau R- (1-\omega)  \bigg[\frac{\sigma_{e}B}{k_{1}+B} +\sigma_{h}I \bigg]S\\ -\alpha S
 \Phi(a,I)I-(\tau +\alpha) R 
\end{bmatrix},$$
$$\Theta(U,V)=\begin{bmatrix}
(1-\omega)  \bigg[\frac{\sigma_{e}B}{k_{1}+B} +\sigma_{h}I\bigg]S - \Phi(a,I)I - \alpha I  \\
\gamma B \bigg(1-\frac{B}{k}\bigg) + \xi I -\beta B -\delta B
\end{bmatrix}.$$}
Now, $\Theta(U,0)=0.$ In order to establish global asymptotic stability, the following assumptions must be satisfied.
i). $U^*$ is globally asymptotically stable for $\frac{dU}{dt}= \Upsilon(U,0)$, ii). $\Theta(U, V)=DV-\hat{\Theta}(U,V); \hat{\Theta}(U,V) \geq 0$ for $(U, V) \in \Re,$ where $D=H_{V}\Theta(U^*, 0)$ is considered as a Metzler matrix (here non-diagonal components are nonnegative) in the region $\Re.$  
From the assumption (I), the model \eqref{eq1} can be represented as 
    $\frac{d}{dt}\begin{pmatrix}
    S\\
    R
    \end{pmatrix}
    =\begin{pmatrix}
    \pi + \tau R-\alpha S\\
    -(\tau + \alpha) R
    \end{pmatrix}.$
Solving the above system, $S(t)=\frac{\pi}{\alpha}$ and $R(t)\rightarrow 0$ as $t \rightarrow \infty.$ Hence, $U^*$ is globally asymptotically stable for $\frac{dU}{dt}=\Upsilon(U, 0)$. Hence, the assumption (I) is satisfied.
Now the matrix $D$ and $\hat{\Theta}(U,V)$ are
$$D=\begin{pmatrix}
 (1-\omega)  \sigma_{h}\frac{\pi}{\alpha} -\phi_{1}-\alpha& (1-\omega)  \sigma_{e}\frac{\pi}{k_{1}\alpha} \\
\xi & \gamma-\beta-\delta
\end{pmatrix}$$
and
$$\hat{\Theta}(U,V)=\begin{pmatrix}
\frac{(1-\omega)\sigma_{e}SB^{2}}{k_{1}(k_{1}+B)}+(\phi_{1}-\phi_{0})I[\frac{a}{I+a}-1]\\
\frac{\gamma B^2}{k}
\end{pmatrix}.$$
In disease-free equilibrium, $\phi_{1}\approx\phi_{0}$. So, we can neglect the term $(\phi_{1}-\phi_{0})I[\frac{a}{I+a}-1]$. Then $\hat{\Theta}(U, V)\geq 0$ for the region $\Re.$ Therefore ,the cholera free equilibrium $W_0$ of the system \eqref{eq1} is considered as globally asymptotically stable in the region $\Re$ when $R_0<1.$ 
\end{proof}
\section{The proof of the Theorem 3}\label{Long-term Dynamics}
\begin{proof}
Assuming  $\pi=\pi^*$ as bifurcation parameter for $R_0=1$, the centre manifold theory is employed for stability analysis at $W^*(S^*, I^*, B^*, R^*)$. Now, the eigenvector corresponding to the eigenvalue zero of the variational matrix of system \eqref{eq1} at $\pi=\pi^*$ is represented by $z= [z_1, z_2, z_3, z_4]^{'}$, where
\begin{align*}
z_1 & = \; \frac{1}{\alpha}\Bigg[\frac{\tau \phi_1}{\tau+\alpha}-\frac{(1-\omega) \sigma_h \pi}{\alpha}-\frac{(1-\omega) \sigma_e \pi \xi}{k_1 \alpha(\beta + \delta -\gamma)}\Bigg] z_2, \\ z_2 & > \;0, \; z_3=\frac{\xi}{\beta + \delta -\gamma}z_2, \;  z_4=\frac{\phi_1}{\tau + \alpha}z_2.
\end{align*}
Similarly, the left eigenvector having zero eigenvalues to the variational matrix  at $\pi=\pi^*$ is given by $w= [w_1, w_2, w_3, w_4]^{'}$, where
   $ w_1=0, w_2>0, w_3=\frac{(\beta+\delta-\gamma)k_1\alpha}{(1-\omega) \sigma_e \pi}w_2, w_4=0.$

We introduce new symbols for the SIBR model as follows:
$S=p_1, I=p_2, B=p_3, R=p_4$ and $\frac{dp_i}{dt}=q_i,$ where $i=1,2,3,4.$ Now, we calculate $q_i$ at $W_0$ and get
\begin{align*}
\frac{\partial^2 q_1}{\partial p_1 \partial p_2} & = \; -{(1-\omega)}{\sigma_h}, \; \frac{\partial^2 q_1}{\partial p_1 \partial p_3} = \; -\frac{1}{k_1}(1-\omega)\sigma_e,\\
\frac{\partial^2 q_1}{\partial p_3\partial p_3} & = \; \frac{2(1-\omega)\sigma_e\pi}{k_1^2\alpha}, \; \frac{\partial^2 q_2}{\partial p_1 \partial p_3} =\frac{(1-\omega)\sigma_e}{k_1}, \\
\frac{\partial^2 q_2}{\partial p_1 \partial p_2} & = \;{(1-\omega)}{\sigma_h}, \; \frac{\partial^2 q_2}{\partial p_2\partial p_2}=\frac{2}{a}(\phi_1-\phi_0), \\
\frac{\partial^2 q_2}{\partial p_1 \partial p_3} & = \; \frac{1}{k_1}(1-\omega)\sigma_e, \; \frac{\partial^2 q_2}{\partial p_3 \partial p_3}=-\frac{2(1-\omega)\sigma_e\pi}{k_1^2\alpha}, \\
\frac{\partial^2 q_3}{\partial p_3 \partial p_3}& = \;-\frac{2\gamma}{k}, \; \frac{\partial^2 q_4}{\partial p_2 \partial p_2}=-\frac{2}{a}(\phi_1-\phi_0).
\end{align*}
The remaining derivatives at $W_0$ become zero. We further calculate the coefficient $\wp$ and $\Im$ based on well-established Theorem 4.1 in Castillo-Chavez et al. \citep{castillo2004dynamical} 
$$\wp=\sum\limits_{i,j,k=1}^{4}w_k z_i z_j\frac{\partial^2q_k(0, \pi^*)}{\partial p_i\partial p_j} \; \text{and}\;  \Im=\sum\limits_{i,k=1}^{4}w_k z_i\frac{\partial^2q_k(0, 0)}{\partial p_i \pi}.$$
Now, we substitute all to determine the coefficient $\wp$ and $\Im$ at threshold $\pi=\pi^*$, we get
   $ \wp=\frac{2(\phi_1-\phi_0)}{\pi^*}w_2z_2z_2-\frac{2(1-\omega)\sigma_e\pi}{k_1^2\alpha}w_2z_3z_3 - \frac{2\gamma}{k}w_3z_3z_3,
    ~\mbox{when}~ \phi_0>\phi_1 ~\mbox{then}~\wp>0$
and
   $ \Im=\frac{(1-\omega)\sigma_h}{\alpha}[w_2w_2 + \frac{w_2z_3}{k_1}]>0.$
Here, the values of $a$ and $b$ indicate negative and positive, respectively. The system \eqref{eq1} experiences forward bifurcation at $R_0=1$. The cholera-present equilibrium $\epsilon^*$ is locally asymptotically stable for $R_0>1.$
\end{proof}

\section{Expression of coefficients }\label{coefficent}
\begin{itemize}{}
    \item[$p_6:$]=$\bigg(\frac{\tau}{\tau+\alpha}\bigg)(1-\omega)\sigma_h{\gamma^3}{\phi_0}-(1-\omega)\bigg[{\gamma^3}{\alpha}+{\gamma^3}{\phi_0}\bigg]$
    \item [$p_5:$]=$\bigg(\frac{\tau}{\tau+\alpha}\bigg)(1-{\omega})\bigg[{\phi_0}A{\xi}k{\gamma^2}+{\sigma_h}{\gamma^3}{k_1}{\phi_0}+A{\xi}k{\sigma_h}{\phi_0}{\gamma^2}\bigg]-(1-\omega)\bigg[{\phi_0}A{\xi}k{\gamma^2}+A{\xi}k{\gamma^2}{\phi_0}+{\gamma^3}{\phi_0}{k_1}+A{\xi}k{\gamma^2}{\phi_0}+A{\xi}k{\gamma^2}{\alpha}+A{\xi}k{\gamma^2}{\alpha}\bigg]$
    \item [$p_4:$]=${\pi}(1-\omega)({\sigma_h}{\gamma^2})\bigg(\frac{\tau}{\tau+\alpha}\bigg)\bigg[{\sigma_h}{\gamma}{A^2}{\xi^2}{k^2}{\phi_0}+{\sigma_h}{k_1}{A}{\xi}{\gamma^2}{\phi_0}+{\phi_0}{\sigma_h}{A}{\xi}{\gamma^2}{k}{k_1}\bigg]-(1-\omega)\bigg[{\phi_0}{A^2}{\xi^2}{k^2}{\gamma}+{\phi_0}{A}{\xi}{k}{k_1}{\gamma^2}+{\phi_0}{A^2}{\xi^2}{k^2}{\gamma}+{A}{\sigma_e}{\phi_o}{\xi^2}{k^2}{\gamma}+{A}{\gamma^2}{\xi}{k}{k_1}{\phi_o} \\+{A^2}{\xi^2}{k^2}{\gamma}{\phi_0}{\sigma_e}{\xi}{k}{\gamma}{\phi_0}+{\sigma_h}{A}{\xi}{k}{k_1}{\phi_0}{\gamma^2}+2{\phi_1}{A}{\xi}k{\gamma^2}+{A^2}{\xi^2}{\alpha}{\gamma}{k}+{A}{\xi}{\alpha}{k}{k_1}{\gamma^2}+{A^2}{\xi^2}{k^2}{\alpha}{\gamma}+{A}{\gamma^2}{\xi}{k}{k_1}{\alpha}+{\sigma_e}{\xi}{k}{\gamma^2}{\alpha}+{A}{\xi}{\alpha}{k}{\gamma^2}{\alpha}+{A}{\xi}{\alpha}{k}{\gamma^2}\bigg]-\bigg[{\gamma^2}{\alpha}{\xi}{k}{\phi_0}+{\gamma^2}{\alpha^2}{\xi}{k}\bigg]$.
    \item [$p_3:$]=$\pi(1-\omega)\bigg[{\sigma_h}A{\xi^2}{k^2}{\gamma}+3{\sigma_h}{\gamma^2}{k_1}+{\sigma_h}A{\xi}k{\gamma}\bigg]\bigg(\frac{\tau}{\tau+\alpha}\bigg)(1-\omega)\bigg[{\sigma_h}{\phi_0}{\gamma}{k_1}{A^2}{\xi^2}{k^2}+{A^3}{\xi^3}{k^3}{\sigma_h}{\phi_0}+{\sigma_e}{\phi_0}A{\xi^2}{k^2}{\gamma}+{\sigma_h}{\phi_0}{A^2}{\xi^2}{k^2}{k_1}{\gamma}+{\phi_0}{\sigma_e}A{\gamma}{\xi^2}{k^2}+{\phi_0}{\sigma_h}{A^2}{\xi^2}{k^2}{k_1}{\gamma}+{\sigma_h}{\phi_1}{A}{a}{\xi^2}{k^2}{\gamma}+{\sigma_h}{\gamma^2}a{\xi}{k}{k_1}{\phi_1}+A{\gamma}{a}{\xi^2}{k^2}{\sigma_h}{\phi_1}\bigg]-(1-\omega)\bigg[{A^2}{\phi_0}{\gamma}{\xi}{k^2}{k_1}{\phi_0}{A^3}{\xi^3}{k^3}+{\phi_0}{\sigma_e}{\xi^2}{A}{k^2}{\gamma}+{\sigma_h}{\phi_0}{A^2}{\xi^2}{k^2}{k_1}{\gamma}+{\sigma_h}{A^2}{\xi^2}{k^2}{k_1}{\gamma}{\phi_0}+{A}{\xi^2}{k^2}{\gamma}{\phi_1}{a}+{\phi_1}{a}{\xi}{k}{k_1}{\gamma^2}+{A}{\xi^2}{k^2}{\gamma}{\phi_1}a +{A^2}{\gamma}{\xi^2}{k^2}{k_1}+{A^3}{\xi^3}{\alpha}{k^3}+{A}{\sigma_e}{\xi^2}{\alpha}{k^2}{\gamma}+{A}{\sigma_e}{\xi^2}{k^2}{\gamma}{\alpha}+{\sigma_h}{A^2}{\xi^2}{k^2}{k_1}{\gamma}{\alpha}+Aa{\xi^2}{k^2}{\gamma}{\alpha}+a{\xi}{\alpha}{k}{k_1}{\gamma^2}+Aa{\xi^2}{k^2}{\alpha}{\gamma}\bigg]-\bigg[{\phi_0}{A}{\xi^2}{\gamma}{\alpha}+{\phi_0}{\gamma^2}{\alpha}{\xi}{k}{k_1}+{A}{\xi^2}{k^2}{\alpha}{\gamma}{\phi_0}+{\gamma^2}{\alpha^2}{\xi}{k}{k_1}+{A}{\xi^2}{k^2}{\alpha^2}{\gamma}+{\gamma}{\alpha^2}{\xi^2}{k^2}{A}\bigg]$
    \item [$p_2:$]=$\pi(1-\omega)\bigg[{\sigma_h}{\gamma}{k_1}{A}{\xi^2}{k^2}+{\sigma_h}{A^2}{\xi^3}{k^3}+{\sigma_e}{\xi}{k}{\gamma}+{\sigma_h}{A}{\xi}{k}{k_1}{\gamma}+{\sigma_h}{a}{\xi^2}{k^2}{\gamma}\bigg]+\bigg(\frac{\tau}{\tau+\alpha}\bigg)(1-\omega)\bigg[{\sigma_e}{\xi}{k}{\phi_0}{A^2}{\xi^2}{k^2}+{\phi_0}{\sigma_h}{A^3}{\xi^3}{k^3}{k_1}+{\sigma_h}{\phi_1}{A}{a}{\xi^2}{k^2}{\gamma}{k_1}+{A}{\xi^2}{k^3}{\sigma_h}{\phi_1}{a}+{\sigma_e}{\phi_1}{\gamma}{a}{\xi^2}{k^2}+{\sigma_h}{\phi_1}{A}{\xi^2}{k^2}{k_1}{\gamma}{a}\bigg]-(1-\omega)\bigg[{\phi_0}{\sigma_e}{A^2}{\xi^3}{k^3}+{\sigma_h}{\phi_0}{A^3}{\xi^3}{k^3}{k_1}+{A}{\gamma}{\xi}{k^2}{k_1}{\phi_1}{a}+{A^2}{\xi^3}{k^3}{a}{\phi_1}+{\sigma_e}{\phi_1}{a}{\xi^2}{k^2}{\gamma}+{\sigma_h}{\phi_1}{A}{a}{\xi^2}{k^2}{k_1}{\gamma}+{A^2}{\sigma_e}{\xi^3}{\alpha}{k^3}+{\sigma_h}{A^3}{\xi^3}{\alpha}{k^3}{k_1}+{\sigma_h}{A}{\xi}{k}{k_1}{\gamma^2}{\alpha}+{A}{a}{\gamma}{\xi^2}{k^2}{\alpha}{k_1}+{A^2}{a}{\xi^3}{k^3}{\alpha}+{\sigma_e}{\xi^2}{k^2}{a}{\alpha}{\gamma}+{\sigma_h}{A}{\xi^2}{k^2}{a}{\alpha}{k_1}{\gamma}\bigg]-\bigg[{\phi_0}{A}{\xi^2}{k^2}{\gamma}{\alpha}{k_1}+{\phi_0}{A^2}{\xi^3}{k^3}{\alpha}+{A}{\alpha}{\xi^2}{k^2}{k_1}{\gamma}{\phi_0}+{\gamma}{\alpha}{a}{\xi^2}{k^2}{\phi_1}+{A}{\alpha^2}{\xi^2}{k^2}{k_1}{\gamma}+{\gamma}{\alpha^2}{A}{\xi^2}{k^2}{k_1}+{A^2}{\xi^3}{k^3}{\alpha^2}\bigg]$
    \item [$p_1:$]=${\pi}(1-\omega)\bigg[{A}{\sigma_e}{\xi^3}{k^3}+{\sigma_h}{A^2}{\xi^3}{k^3}{k_1}+{a}{\xi^2}{k^2}{\sigma_h}{\gamma}{k_1}\bigg]+\bigg(\frac{\tau}{\tau+\alpha}\bigg)(1-\omega)\bigg[{\sigma_e}{\phi_1}{A}{a}{\xi^3}{k^3}+{\sigma_h}{\phi_1}{A^2}{a}{\xi^3}{k^3}{k_1}\bigg]-(1-\omega)\bigg[{A}{\sigma_e}{\xi^3}{k^3}{\phi_1}{a}+{\phi_1}{\sigma_h}{A^2}{\xi^3}{k^3}{a}{k_1}+{\sigma_h}{A^2}{\xi^2}{k^2}{k_1}{\gamma}{\alpha}+{A}{\sigma_e}{a}{\xi^3}{\alpha}{k^3}+{\sigma_h}{A^2}{\xi^3}{k^3}{k_1}\bigg]-\bigg[{\phi_0}{A^2}{\alpha}{\xi^3}{k^3}{k_1}+{\gamma}{\alpha}{\xi^2}{k^2}{k_1}{a}{\phi_1}+{A}{a}{\xi^3}{k^3}{\alpha}{\phi_1}+{A^2}{\alpha^2}{\xi^3}{k^3}{k_1}+{A}{\xi^3}{k^3}{a}{\alpha^2}\bigg]$
    \item [$p_0:$]=$\pi(1-\omega)\bigg[{a}{\xi^2}{k^3}{\sigma_e}{\xi}+{A}{\sigma_h}{a}{\xi^3}{k^3}{k_1}+{\sigma_h}{A}{\xi^3}{a}{k^3}{k_1}+{\sigma_h}{A}{\xi^3}{k^3}{a}\bigg]-\bigg[{A}{\alpha^2}{\xi^3}{k^3}{k_1}+{\gamma}{\alpha^2}{\xi^2}{k^2}{a}\bigg]$
\end{itemize}




\printcredits

\section*{Data and code availability}
The reported cases are collected from the Health Ministry of Malawi: Cholera National Information Dashboard, available at \href{https://cholera.health.gov.mw/surveillance}{Malawi}. For the reproducibility of our work, we have uploaded the data and codes used in this study in \url{https://github.com/ctanujit/EIML}.
\section*{Declaration of competing interest}
The authors declare no conflict of interest.
\section*{Acknowledgments} Adrita Ghosh, junior research fellow, is supported by the University Grant Commission (UGC), India, and is also grateful to UGC for the support.

\bibliographystyle{cas-model2-names}

\bibliography{Cholera_ref}

\begin{thebibliography}{62}
\expandafter\ifx\csname natexlab\endcsname\relax\def\natexlab#1{#1}\fi
\providecommand{\url}[1]{\texttt{#1}}
\providecommand{\href}[2]{#2}
\providecommand{\path}[1]{#1}
\providecommand{\DOIprefix}{doi:}
\providecommand{\ArXivprefix}{arXiv:}
\providecommand{\URLprefix}{URL: }
\providecommand{\Pubmedprefix}{pmid:}
\providecommand{\doi}[1]{\href{http://dx.doi.org/#1}{\path{#1}}}
\providecommand{\Pubmed}[1]{\href{pmid:#1}{\path{#1}}}
\providecommand{\bibinfo}[2]{#2}
\ifx\xfnm\relax \def\xfnm[#1]{\unskip,\space#1}\fi
\bibitem[{Acharya et~al.(2024)Acharya, Mondal, Upadhyay and Das}]{Acharya2024}
\bibinfo{author}{Acharya, S.}, \bibinfo{author}{Mondal, B.}, \bibinfo{author}{Upadhyay, R.K.}, \bibinfo{author}{Das, P.}, \bibinfo{year}{2024}.
\newblock \bibinfo{title}{Exploring noise-induced dynamics and optimal control strategy of isir cholera transmission model}.
\newblock \bibinfo{journal}{Nonlinear Dynamics} \bibinfo{volume}{112}, \bibinfo{pages}{3951–3975}.
\bibitem[{Assimakopoulos and Nikolopoulos(2000)}]{assimakopoulos2000theta}
\bibinfo{author}{Assimakopoulos, V.}, \bibinfo{author}{Nikolopoulos, K.}, \bibinfo{year}{2000}.
\newblock \bibinfo{title}{The theta model: a decomposition approach to forecasting}.
\newblock \bibinfo{journal}{International Journal of Forecasting} \bibinfo{volume}{16}, \bibinfo{pages}{521--530}.
\bibitem[{Barman et~al.(2025)Barman, Panja, Mishra and Chakraborty}]{barman2025epidemic}
\bibinfo{author}{Barman, M.}, \bibinfo{author}{Panja, M.}, \bibinfo{author}{Mishra, N.}, \bibinfo{author}{Chakraborty, T.}, \bibinfo{year}{2025}.
\newblock \bibinfo{title}{Epidemic-guided deep learning for spatiotemporal forecasting of tuberculosis outbreak}.
\newblock \bibinfo{journal}{arXiv preprint arXiv:2502.10786} .
\bibitem[{Batmanova et~al.(2022)Batmanova, Kuc, Maksimenko, Savosenkov, Grigorev, Gordleeva, Kazantsev, Korchagin and Hramov}]{eeg}
\bibinfo{author}{Batmanova, A.}, \bibinfo{author}{Kuc, A.}, \bibinfo{author}{Maksimenko, V.}, \bibinfo{author}{Savosenkov, A.}, \bibinfo{author}{Grigorev, N.}, \bibinfo{author}{Gordleeva, S.}, \bibinfo{author}{Kazantsev, V.}, \bibinfo{author}{Korchagin, S.}, \bibinfo{author}{Hramov, A.E.}, \bibinfo{year}{2022}.
\newblock \bibinfo{title}{Predicting perceptual decision-making errors using eeg and machine learning}.
\newblock \bibinfo{journal}{Mathematics} \bibinfo{volume}{10}, \bibinfo{pages}{3153}.
\bibitem[{Box(2013)}]{box2013box}
\bibinfo{author}{Box, G.}, \bibinfo{year}{2013}.
\newblock \bibinfo{title}{Box and jenkins: time series analysis, forecasting and control}, in: \bibinfo{booktitle}{A Very British Affair: Six Britons and the Development of Time Series Analysis During the 20th Century}. \bibinfo{publisher}{Springer}, pp. \bibinfo{pages}{161--215}.
\bibitem[{Camacho et~al.(2018)Camacho, Bouhenia, Alyusfi and Alkohlani}]{CAMACHO2018e680}
\bibinfo{author}{Camacho, A.}, \bibinfo{author}{Bouhenia, M.}, \bibinfo{author}{Alyusfi, R.}, \bibinfo{author}{Alkohlani, A.}, \bibinfo{year}{2018}.
\newblock \bibinfo{title}{Cholera epidemic in yemen, 2016–18: an analysis of surveillance data}.
\newblock \bibinfo{journal}{The Lancet Global Health} \bibinfo{volume}{6}, \bibinfo{pages}{e680--e690}.
\bibitem[{Castillo-Chavez and Song(2004)}]{castillo2004dynamical}
\bibinfo{author}{Castillo-Chavez, C.}, \bibinfo{author}{Song, B.}, \bibinfo{year}{2004}.
\newblock \bibinfo{title}{Dynamical models of tuberculosis and their applications}.
\newblock \bibinfo{journal}{Mathematical Biosciences \& Engineering} \bibinfo{volume}{1}, \bibinfo{pages}{361--404}.
\bibitem[{Chakraborty et~al.(2019)Chakraborty, Chattopadhyay and Ghosh}]{chakraborty2019forecasting}
\bibinfo{author}{Chakraborty, T.}, \bibinfo{author}{Chattopadhyay, S.}, \bibinfo{author}{Ghosh, I.}, \bibinfo{year}{2019}.
\newblock \bibinfo{title}{Forecasting dengue epidemics using a hybrid methodology}.
\newblock \bibinfo{journal}{Physica A: Statistical Mechanics and its Applications} \bibinfo{volume}{527}, \bibinfo{pages}{121266}.
\bibitem[{Constantin(2010)}]{z90874}
\bibinfo{author}{Constantin, A.}, \bibinfo{year}{2010}.
\newblock \bibinfo{title}{On {Nagumo}'s theorem}.
\newblock \bibinfo{journal}{Proc. Japan Acad., Ser. A} \bibinfo{volume}{86}, \bibinfo{pages}{41--44}.
\bibitem[{Daisy et~al.(2020)Daisy, Saiful~Islam, Akanda, Faruque, Amin and Jensen}]{daisy2020developing}
\bibinfo{author}{Daisy, S.S.}, \bibinfo{author}{Saiful~Islam, A.}, \bibinfo{author}{Akanda, A.S.}, \bibinfo{author}{Faruque, A.S.G.}, \bibinfo{author}{Amin, N.}, \bibinfo{author}{Jensen, P.K.M.}, \bibinfo{year}{2020}.
\newblock \bibinfo{title}{Developing a forecasting model for cholera incidence in dhaka megacity through time series climate data}.
\newblock \bibinfo{journal}{Journal of Water and Health} \bibinfo{volume}{18}, \bibinfo{pages}{207--223}.
\bibitem[{Das et~al.(2021a)Das, Nadim, Das and Das}]{Das20021}
\bibinfo{author}{Das, P.}, \bibinfo{author}{Nadim, S.S.}, \bibinfo{author}{Das, S.}, \bibinfo{author}{Das, P.}, \bibinfo{year}{2021}a.
\newblock \bibinfo{title}{Dynamics of covid-19 transmission with comorbidity: a data driven modelling based approach}.
\newblock \bibinfo{journal}{Nonlinear Dynamics} \bibinfo{volume}{106}, \bibinfo{pages}{1197--1211}.
\bibitem[{Das et~al.(2021b)Das, Upadhyay, Misra, Rihan, Das and Ghosh}]{Das20201}
\bibinfo{author}{Das, P.}, \bibinfo{author}{Upadhyay, R.K.}, \bibinfo{author}{Misra, A.K.}, \bibinfo{author}{Rihan, F.A.}, \bibinfo{author}{Das, P.}, \bibinfo{author}{Ghosh, D.}, \bibinfo{year}{2021}b.
\newblock \bibinfo{title}{Mathematical model of covid-19 with comorbidity and controlling using non-pharmaceutical interventions and vaccination}.
\newblock \bibinfo{journal}{Nonlinear Dynamics} \bibinfo{volume}{106}, \bibinfo{pages}{1213--1227}.
\bibitem[{De~Livera et~al.(2011)De~Livera, Hyndman and Snyder}]{de2011forecasting}
\bibinfo{author}{De~Livera, A.M.}, \bibinfo{author}{Hyndman, R.J.}, \bibinfo{author}{Snyder, R.D.}, \bibinfo{year}{2011}.
\newblock \bibinfo{title}{Forecasting time series with complex seasonal patterns using exponential smoothing}.
\newblock \bibinfo{journal}{Journal of the American Statistical Association} \bibinfo{volume}{106}, \bibinfo{pages}{1513--1527}.
\bibitem[{Entorf(1997)}]{entorf1997random}
\bibinfo{author}{Entorf, H.}, \bibinfo{year}{1997}.
\newblock \bibinfo{title}{Random walks with drifts: Nonsense regression and spurious fixed-effect estimation}.
\newblock \bibinfo{journal}{Journal of Econometrics} \bibinfo{volume}{80}, \bibinfo{pages}{287--296}.
\bibitem[{Faraway and Chatfield(1998)}]{faraway1998time}
\bibinfo{author}{Faraway, J.}, \bibinfo{author}{Chatfield, C.}, \bibinfo{year}{1998}.
\newblock \bibinfo{title}{Time series forecasting with neural networks: a comparative study using the air line data}.
\newblock \bibinfo{journal}{Journal of the Royal Statistical Society Series C: Applied Statistics} \bibinfo{volume}{47}, \bibinfo{pages}{231--250}.
\bibitem[{Faruque et~al.(2005)Faruque, Islam, Ahmad, Faruque, Sack, Nair and Mekalanos}]{pnas05020}
\bibinfo{author}{Faruque, S.M.}, \bibinfo{author}{Islam, M.J.}, \bibinfo{author}{Ahmad, Q.S.}, \bibinfo{author}{Faruque, A.S.G.}, \bibinfo{author}{Sack, D.A.}, \bibinfo{author}{Nair, G.B.}, \bibinfo{author}{Mekalanos, J.J.}, \bibinfo{year}{2005}.
\newblock \bibinfo{title}{Self-limiting nature of seasonal cholera epidemics: Role of host-mediated amplification of phage}.
\newblock \bibinfo{journal}{Proceedings of the National Academy of Sciences} \bibinfo{volume}{102}, \bibinfo{pages}{6119--6124}.
\bibitem[{Funk et~al.(2010)Funk, Salathé and Jansen}]{human_behaviour}
\bibinfo{author}{Funk, S.}, \bibinfo{author}{Salathé, M.}, \bibinfo{author}{Jansen, V.A.A.}, \bibinfo{year}{2010}.
\newblock \bibinfo{title}{Modelling the influence of human behaviour on the spread of infectious diseases: a review}.
\newblock \bibinfo{journal}{Journal of The Royal Society Interface} \bibinfo{volume}{7}, \bibinfo{pages}{1247--1256}.
\bibitem[{Ghosh and Chakraborty(2021)}]{ghosh2021integrated}
\bibinfo{author}{Ghosh, I.}, \bibinfo{author}{Chakraborty, T.}, \bibinfo{year}{2021}.
\newblock \bibinfo{title}{An integrated deterministic--stochastic approach for forecasting the long-term trajectories of covid-19}.
\newblock \bibinfo{journal}{International Journal of Modeling, Simulation, and Scientific Computing} \bibinfo{volume}{12}, \bibinfo{pages}{2141001}.
\bibitem[{Ghosh et~al.(2024)Ghosh, Roy, Perc and Ghosh}]{GHOSH}
\bibinfo{author}{Ghosh, S.}, \bibinfo{author}{Roy, S.}, \bibinfo{author}{Perc, M.}, \bibinfo{author}{Ghosh, D.}, \bibinfo{year}{2024}.
\newblock \bibinfo{title}{The eco-evolutionary dynamics of two strategic species: From the predator-prey to the innocent-spreader rumor model}.
\newblock \bibinfo{journal}{Journal of Theoretical Biology} \bibinfo{volume}{595}, \bibinfo{pages}{111955}.
\bibitem[{Haario et~al.(2006)Haario, Laine, Mira and Saksman}]{haario2006dram}
\bibinfo{author}{Haario, H.}, \bibinfo{author}{Laine, M.}, \bibinfo{author}{Mira, A.}, \bibinfo{author}{Saksman, E.}, \bibinfo{year}{2006}.
\newblock \bibinfo{title}{Dram: efficient adaptive mcmc}.
\newblock \bibinfo{journal}{Statistics and Computing} \bibinfo{volume}{16}, \bibinfo{pages}{339--354}.
\bibitem[{Hartley et~al.(2006)Hartley, Morris and Smith}]{pmed}
\bibinfo{author}{Hartley, D.M.}, \bibinfo{author}{Morris, Jr, J.G.}, \bibinfo{author}{Smith, D.L.}, \bibinfo{year}{2006}.
\newblock \bibinfo{title}{Hyperinfectivity: A critical element in the ability of v. cholerae to cause epidemics?}
\newblock \bibinfo{journal}{PLOS Medicine} \bibinfo{volume}{3}, \bibinfo{pages}{e7}.
\bibitem[{Hazelbag et~al.(2020)Hazelbag, Dushoff, Dominic, Mthombothi and Delva}]{hazelbag2020calibration}
\bibinfo{author}{Hazelbag, C.M.}, \bibinfo{author}{Dushoff, J.}, \bibinfo{author}{Dominic, E.M.}, \bibinfo{author}{Mthombothi, Z.E.}, \bibinfo{author}{Delva, W.}, \bibinfo{year}{2020}.
\newblock \bibinfo{title}{Calibration of individual-based models to epidemiological data: A systematic review}.
\newblock \bibinfo{journal}{PLoS Computational Biology} \bibinfo{volume}{16}, \bibinfo{pages}{e1007893}.
\bibitem[{Hyndman et~al.(2008)Hyndman, Koehler, Ord and Snyder}]{hyndman2008forecasting}
\bibinfo{author}{Hyndman, R.}, \bibinfo{author}{Koehler, A.B.}, \bibinfo{author}{Ord, J.K.}, \bibinfo{author}{Snyder, R.D.}, \bibinfo{year}{2008}.
\newblock \bibinfo{title}{Forecasting with exponential smoothing: the state space approach}.
\newblock \bibinfo{publisher}{Springer Science \& Business Media}.
\bibitem[{Hyndman and Athanasopoulos(2018)}]{hyndman2018forecasting}
\bibinfo{author}{Hyndman, R.J.}, \bibinfo{author}{Athanasopoulos, G.}, \bibinfo{year}{2018}.
\newblock \bibinfo{title}{Forecasting: principles and practice}.
\newblock \bibinfo{publisher}{OTexts}.
\bibitem[{K.B. et~al.(2025)K.B., Jose, Jirawattanapanit and Mathew}]{KB2025111988}
\bibinfo{author}{K.B., H.M.}, \bibinfo{author}{Jose, S.A.}, \bibinfo{author}{Jirawattanapanit, A.}, \bibinfo{author}{Mathew, K.}, \bibinfo{year}{2025}.
\newblock \bibinfo{title}{A comprehensive study on tuberculosis prediction models: Integrating machine learning into epidemiological analysis}.
\newblock \bibinfo{journal}{Journal of Theoretical Biology} \bibinfo{volume}{597}, \bibinfo{pages}{111988}.
\bibitem[{Kharazmi et~al.(2021)Kharazmi, Cai, Zheng, Zhang, Lin and Karniadakis}]{Kharazmi2021}
\bibinfo{author}{Kharazmi, E.}, \bibinfo{author}{Cai, M.}, \bibinfo{author}{Zheng, X.}, \bibinfo{author}{Zhang, Z.}, \bibinfo{author}{Lin, G.}, \bibinfo{author}{Karniadakis, G.E.}, \bibinfo{year}{2021}.
\newblock \bibinfo{title}{Identifiability and predictability of integer- and fractional-order epidemiological models using physics-informed neural networks}.
\newblock \bibinfo{journal}{Nature Computational Science} \bibinfo{volume}{1}, \bibinfo{pages}{744--753}.
\bibitem[{King et~al.(2008)King, Ionides, Pascual and Bouma}]{King20008}
\bibinfo{author}{King, A.A.}, \bibinfo{author}{Ionides, E.L.}, \bibinfo{author}{Pascual, M.}, \bibinfo{author}{Bouma, M.J.}, \bibinfo{year}{2008}.
\newblock \bibinfo{title}{Inapparent infections and cholera dynamics}.
\newblock \bibinfo{journal}{Nature} \bibinfo{volume}{454}, \bibinfo{pages}{877--880}.
\bibitem[{Kong et~al.(2014)Kong, Davis and Wang}]{Kong20014}
\bibinfo{author}{Kong, J.D.}, \bibinfo{author}{Davis, W.}, \bibinfo{author}{Wang, H.}, \bibinfo{year}{2014}.
\newblock \bibinfo{title}{Dynamics of a cholera transmission model with immunological threshold and natural phage control in reservoir}.
\newblock \bibinfo{journal}{Bulletin of Mathematical Biology} \bibinfo{volume}{76}, \bibinfo{pages}{2025--2051}.
\bibitem[{Leo et~al.(2019)Leo, Luhanga, Michael et~al.}]{leo2019machine}
\bibinfo{author}{Leo, J.}, \bibinfo{author}{Luhanga, E.}, \bibinfo{author}{Michael, K.}, et~al., \bibinfo{year}{2019}.
\newblock \bibinfo{title}{Machine learning model for imbalanced cholera dataset in tanzania}.
\newblock \bibinfo{journal}{The Scientific World Journal} \bibinfo{volume}{2019}.
\bibitem[{Lopez et~al.(2020)Lopez, Dutta, Qadri, Sovann, Pandey, Hamzah, Memon, Iamsirithaworn, Dang, Chowdhury et~al.}]{Lopez2020c}
\bibinfo{author}{Lopez, A.L.}, \bibinfo{author}{Dutta, S.}, \bibinfo{author}{Qadri, F.}, \bibinfo{author}{Sovann, L.}, \bibinfo{author}{Pandey, B.D.}, \bibinfo{author}{Hamzah, W.M.B.}, \bibinfo{author}{Memon, I.}, \bibinfo{author}{Iamsirithaworn, S.}, \bibinfo{author}{Dang, D.A.}, \bibinfo{author}{Chowdhury, F.}, et~al., \bibinfo{year}{2020}.
\newblock \bibinfo{title}{Cholera in selected countries in asia}.
\newblock \bibinfo{journal}{Vaccine} \bibinfo{volume}{38}, \bibinfo{pages}{A18--A24}.
\bibitem[{Luby et~al.(2020)Luby, Davis, Brown, Gorelick and Wong}]{LUBY2020A110}
\bibinfo{author}{Luby, S.P.}, \bibinfo{author}{Davis, J.}, \bibinfo{author}{Brown, R.R.}, \bibinfo{author}{Gorelick, S.M.}, \bibinfo{author}{Wong, T.H.}, \bibinfo{year}{2020}.
\newblock \bibinfo{title}{Broad approaches to cholera control in asia: Water, sanitation and handwashing}.
\newblock \bibinfo{journal}{Vaccine} \bibinfo{volume}{38}, \bibinfo{pages}{A110--A117}.
\bibitem[{Maity et~al.(2023)Maity, Saha, Ghosh and Chattopadhyay}]{Maity20123}
\bibinfo{author}{Maity, B.}, \bibinfo{author}{Saha, B.}, \bibinfo{author}{Ghosh, I.}, \bibinfo{author}{Chattopadhyay, J.}, \bibinfo{year}{2023}.
\newblock \bibinfo{title}{Model-based estimation of expected time to cholera extinction in lusaka, zambia}.
\newblock \bibinfo{journal}{Bulletin of Mathematical Biology} \bibinfo{volume}{85}, \bibinfo{pages}{55}.
\bibitem[{Mari et~al.(2015)Mari, Bertuzzo, Finger and Casagrandi}]{do88}
\bibinfo{author}{Mari, L.}, \bibinfo{author}{Bertuzzo, E.}, \bibinfo{author}{Finger, F.}, \bibinfo{author}{Casagrandi, R.}, \bibinfo{year}{2015}.
\newblock \bibinfo{title}{On the predictive ability of mechanistic models for the haitian cholera epidemic}.
\newblock \bibinfo{journal}{Journal of The Royal Society Interface} \bibinfo{volume}{12}, \bibinfo{pages}{20140840}.
\bibitem[{Mari et~al.(2012)Mari, Bertuzzo, Righetto, Casagrandi and Gatto}]{1098304}
\bibinfo{author}{Mari, L.}, \bibinfo{author}{Bertuzzo, E.}, \bibinfo{author}{Righetto, L.}, \bibinfo{author}{Casagrandi, R.}, \bibinfo{author}{Gatto, M.}, \bibinfo{year}{2012}.
\newblock \bibinfo{title}{Modelling cholera epidemics: the role of waterways, human mobility and sanitation}.
\newblock \bibinfo{journal}{Journal of The Royal Society Interface} \bibinfo{volume}{9}, \bibinfo{pages}{376--388}.
\bibitem[{Marino et~al.(2008)Marino, Hogue, Ray and Kirschner}]{marino2008methodology}
\bibinfo{author}{Marino, S.}, \bibinfo{author}{Hogue, I.B.}, \bibinfo{author}{Ray, C.J.}, \bibinfo{author}{Kirschner, D.E.}, \bibinfo{year}{2008}.
\newblock \bibinfo{title}{A methodology for performing global uncertainty and sensitivity analysis in systems biology}.
\newblock \bibinfo{journal}{Journal of Theoretical Biology} \bibinfo{volume}{254}, \bibinfo{pages}{178--196}.
\bibitem[{Martinez et~al.(2017)Martinez, Reiner~Jr, Cash, Rod{\'o}, Shahjahan~Mondal, Roy, Yunus, Faruque, Huq, King et~al.}]{martinez2017cholera}
\bibinfo{author}{Martinez, P.P.}, \bibinfo{author}{Reiner~Jr, R.C.}, \bibinfo{author}{Cash, B.A.}, \bibinfo{author}{Rod{\'o}, X.}, \bibinfo{author}{Shahjahan~Mondal, M.}, \bibinfo{author}{Roy, M.}, \bibinfo{author}{Yunus, M.}, \bibinfo{author}{Faruque, A.S.}, \bibinfo{author}{Huq, S.}, \bibinfo{author}{King, A.A.}, et~al., \bibinfo{year}{2017}.
\newblock \bibinfo{title}{Cholera forecast for dhaka, bangladesh, with the 2015-2016 el ni{\~n}o: lessons learned}.
\newblock \bibinfo{journal}{PloS one} \bibinfo{volume}{12}, \bibinfo{pages}{e0172355}.
\bibitem[{Meszaros et~al.(2020)Meszaros, Miller-Dickson, Baffour-Awuah, Almagro~Moreno and Ogbunugafor}]{Meszaros2020a}
\bibinfo{author}{Meszaros, V.A.}, \bibinfo{author}{Miller-Dickson, M.D.}, \bibinfo{author}{Baffour-Awuah, Junior, F.}, \bibinfo{author}{Almagro~Moreno, S.}, \bibinfo{author}{Ogbunugafor, C.B.}, \bibinfo{year}{2020}.
\newblock \bibinfo{title}{Direct transmission via households informs models of disease and intervention dynamics in cholera}.
\newblock \bibinfo{journal}{PLoS One} \bibinfo{volume}{15}, \bibinfo{pages}{e0229837}.
\bibitem[{Miggo et~al.(2023)Miggo, Harawa, Kangwerema, Knovicks, Mfune, Safari, Kaunda, Kalua, Sefu, Phiri and Patel}]{Migg0o2023}
\bibinfo{author}{Miggo, M.}, \bibinfo{author}{Harawa, G.}, \bibinfo{author}{Kangwerema, A.}, \bibinfo{author}{Knovicks, S.}, \bibinfo{author}{Mfune, C.}, \bibinfo{author}{Safari, J.}, \bibinfo{author}{Kaunda, J.T.}, \bibinfo{author}{Kalua, J.}, \bibinfo{author}{Sefu, G.}, \bibinfo{author}{Phiri, E.}, \bibinfo{author}{Patel, P.}, \bibinfo{year}{2023}.
\newblock \bibinfo{title}{Fight against cholera outbreak, efforts and challenges in malawi}.
\newblock \bibinfo{journal}{Health Sci. Rep.} \bibinfo{volume}{6}, \bibinfo{pages}{e1594}.
\bibitem[{Miller~Neilan et~al.(2010)Miller~Neilan, Schaefer, Gaff, Fister and Lenhart}]{Miller2010}
\bibinfo{author}{Miller~Neilan, R.L.}, \bibinfo{author}{Schaefer, E.}, \bibinfo{author}{Gaff, H.}, \bibinfo{author}{Fister, K.R.}, \bibinfo{author}{Lenhart, S.}, \bibinfo{year}{2010}.
\newblock \bibinfo{title}{Modeling optimal intervention strategies for cholera}.
\newblock \bibinfo{journal}{Bulletin of Mathematical Biology} \bibinfo{volume}{72}, \bibinfo{pages}{2004--2018}.
\bibitem[{Montero et~al.(2023a)Montero, Vidal and Velasco}]{Montero2023-yc}
\bibinfo{author}{Montero, D.A.}, \bibinfo{author}{Vidal, R.M.}, \bibinfo{author}{Velasco, J.}, \bibinfo{year}{2023}a.
\newblock \bibinfo{title}{Vibrio cholerae, classification, pathogenesis, immune response, and trends in vaccine development}.
\newblock \bibinfo{journal}{Front. Med. (Lausanne)} \bibinfo{volume}{10}, \bibinfo{pages}{1155751}.
\bibitem[{Montero et~al.(2023b)Montero, Vidal, Velasco, George and Lucero}]{Montero2023}
\bibinfo{author}{Montero, D.A.}, \bibinfo{author}{Vidal, R.M.}, \bibinfo{author}{Velasco, J.}, \bibinfo{author}{George, S.}, \bibinfo{author}{Lucero, Y.}, \bibinfo{year}{2023}b.
\newblock \bibinfo{title}{Vibrio cholerae, classification, pathogenesis, immune response, and trends in vaccine development}.
\newblock \bibinfo{journal}{Front. Med. (Lausanne)} \bibinfo{volume}{10}, \bibinfo{pages}{1155751}.
\bibitem[{Moore and Lessler(2015)}]{Moore20150}
\bibinfo{author}{Moore, S.M.}, \bibinfo{author}{Lessler, J.}, \bibinfo{year}{2015}.
\newblock \bibinfo{title}{Optimal allocation of the limited oral cholera vaccine supply between endemic and epidemic settings}.
\newblock \bibinfo{journal}{J. R. Soc. Interface} \bibinfo{volume}{12}, \bibinfo{pages}{20150703}.
\bibitem[{Morens et~al.(2004)Morens, Folkers and Fauci}]{Morens20004}
\bibinfo{author}{Morens, D.M.}, \bibinfo{author}{Folkers, G.K.}, \bibinfo{author}{Fauci, A.S.}, \bibinfo{year}{2004}.
\newblock \bibinfo{title}{The challenge of emerging and re-emerging infectious diseases}.
\newblock \bibinfo{journal}{Nature} \bibinfo{volume}{430}, \bibinfo{pages}{242--249}.
\bibitem[{Mukandavire et~al.(2011)Mukandavire, Liao, Wang, Gaff, Smith and Morris}]{pnas0008}
\bibinfo{author}{Mukandavire, Z.}, \bibinfo{author}{Liao, S.}, \bibinfo{author}{Wang, J.}, \bibinfo{author}{Gaff, H.}, \bibinfo{author}{Smith, D.L.}, \bibinfo{author}{Morris, J.G.}, \bibinfo{year}{2011}.
\newblock \bibinfo{title}{Estimating the reproductive numbers for the 2008–2009 cholera outbreaks in zimbabwe}.
\newblock \bibinfo{journal}{Proceedings of the National Academy of Sciences} \bibinfo{volume}{108}, \bibinfo{pages}{8767--8772}.
\bibitem[{Mukandavire and Morris(2015)}]{Mukandavire20}
\bibinfo{author}{Mukandavire, Z.}, \bibinfo{author}{Morris, Jr, J.G.}, \bibinfo{year}{2015}.
\newblock \bibinfo{title}{Modeling the epidemiology of cholera to prevent disease transmission in developing countries}.
\newblock \bibinfo{journal}{Microbiol. Spectr.} \bibinfo{volume}{3}.
\bibitem[{Muzembo et~al.(2022)Muzembo, Kitahara, Debnath, Ohno, Okamoto and Miyoshi}]{Muzembo2022}
\bibinfo{author}{Muzembo, B.A.}, \bibinfo{author}{Kitahara, K.}, \bibinfo{author}{Debnath, A.}, \bibinfo{author}{Ohno, A.}, \bibinfo{author}{Okamoto, K.}, \bibinfo{author}{Miyoshi, S.I.}, \bibinfo{year}{2022}.
\newblock \bibinfo{title}{Cholera outbreaks in india, 2011-2020: A systematic review}.
\newblock \bibinfo{journal}{Int. J. Environ. Res. Public Health} \bibinfo{volume}{19}, \bibinfo{pages}{5738}.
\bibitem[{Nelson et~al.(2009)Nelson, Harris and Morris}]{Nelson20009}
\bibinfo{author}{Nelson, E.J.}, \bibinfo{author}{Harris, J.B.}, \bibinfo{author}{Morris, Jr, J.G.}, \bibinfo{year}{2009}.
\newblock \bibinfo{title}{Cholera transmission: the host, pathogen and bacteriophage dynamic}.
\newblock \bibinfo{journal}{Nat. Rev. Microbiol.} \bibinfo{volume}{7}, \bibinfo{pages}{693--702}.
\bibitem[{Ning et~al.(2023a)Ning, Guan, Li, Wei and Chen}]{v15081749}
\bibinfo{author}{Ning, X.}, \bibinfo{author}{Guan, J.}, \bibinfo{author}{Li, X.A.}, \bibinfo{author}{Wei, Y.}, \bibinfo{author}{Chen, F.}, \bibinfo{year}{2023}a.
\newblock \bibinfo{title}{Physics-informed neural networks integrating compartmental model for analyzing covid-19 transmission dynamics}.
\newblock \bibinfo{journal}{Viruses} \bibinfo{volume}{15}.
\bibitem[{Ning et~al.(2023b)Ning, Jia, Wei, Li and Chen}]{NING2023106693}
\bibinfo{author}{Ning, X.}, \bibinfo{author}{Jia, L.}, \bibinfo{author}{Wei, Y.}, \bibinfo{author}{Li, X.A.}, \bibinfo{author}{Chen, F.}, \bibinfo{year}{2023}b.
\newblock \bibinfo{title}{Epi-dnns: Epidemiological priors informed deep neural networks for modeling covid-19 dynamics}.
\newblock \bibinfo{journal}{Computers in Biology and Medicine} \bibinfo{volume}{158}, \bibinfo{pages}{106693}.
\bibitem[{Nishiura et~al.(2017)Nishiura, Tsuzuki, Yuan, Yamaguchi and Asai}]{nishiura2017transmission}
\bibinfo{author}{Nishiura, H.}, \bibinfo{author}{Tsuzuki, S.}, \bibinfo{author}{Yuan, B.}, \bibinfo{author}{Yamaguchi, T.}, \bibinfo{author}{Asai, Y.}, \bibinfo{year}{2017}.
\newblock \bibinfo{title}{Transmission dynamics of cholera in yemen, 2017: a real time forecasting}.
\newblock \bibinfo{journal}{Theoretical Biology and Medical Modelling} \bibinfo{volume}{14}, \bibinfo{pages}{1--8}.
\bibitem[{Nyabadza et~al.(2019)Nyabadza, Aduamah and Mushanyu}]{Nyabadza20019}
\bibinfo{author}{Nyabadza, F.}, \bibinfo{author}{Aduamah, J.M.}, \bibinfo{author}{Mushanyu, J.}, \bibinfo{year}{2019}.
\newblock \bibinfo{title}{Modelling cholera transmission dynamics in the presence of limited resources}.
\newblock \bibinfo{journal}{BMC Research Notes} \bibinfo{volume}{12}, \bibinfo{pages}{475}.
\bibitem[{Panja et~al.(2023a)Panja, Chakraborty, Kumar and Liu}]{panja2023epicasting}
\bibinfo{author}{Panja, M.}, \bibinfo{author}{Chakraborty, T.}, \bibinfo{author}{Kumar, U.}, \bibinfo{author}{Liu, N.}, \bibinfo{year}{2023}a.
\newblock \bibinfo{title}{Epicasting: an ensemble wavelet neural network for forecasting epidemics}.
\newblock \bibinfo{journal}{Neural Networks} \bibinfo{volume}{165}, \bibinfo{pages}{185--212}.
\bibitem[{Panja et~al.(2023b)Panja, Chakraborty, Nadim, Ghosh, Kumar and Liu}]{panja2023ensemble}
\bibinfo{author}{Panja, M.}, \bibinfo{author}{Chakraborty, T.}, \bibinfo{author}{Nadim, S.S.}, \bibinfo{author}{Ghosh, I.}, \bibinfo{author}{Kumar, U.}, \bibinfo{author}{Liu, N.}, \bibinfo{year}{2023}b.
\newblock \bibinfo{title}{An ensemble neural network approach to forecast dengue outbreak based on climatic condition}.
\newblock \bibinfo{journal}{Chaos, Solitons \& Fractals} \bibinfo{volume}{167}, \bibinfo{pages}{113124}.
\bibitem[{Pascual et~al.(2000)Pascual, Rodó, Ellner, Colwell and Bouma}]{nino}
\bibinfo{author}{Pascual, M.}, \bibinfo{author}{Rodó, X.}, \bibinfo{author}{Ellner, S.P.}, \bibinfo{author}{Colwell, R.}, \bibinfo{author}{Bouma, M.J.}, \bibinfo{year}{2000}.
\newblock \bibinfo{title}{Cholera dynamics and el niño-southern oscillation}.
\newblock \bibinfo{journal}{Science} \bibinfo{volume}{289}, \bibinfo{pages}{1766--1769}.
\bibitem[{Pasetto et~al.(2018)Pasetto, Finger, Camacho, Grandesso, Cohuet, Lemaitre, Azman, Luquero, Bertuzzo and Rinaldo}]{pasetto2018near}
\bibinfo{author}{Pasetto, D.}, \bibinfo{author}{Finger, F.}, \bibinfo{author}{Camacho, A.}, \bibinfo{author}{Grandesso, F.}, \bibinfo{author}{Cohuet, S.}, \bibinfo{author}{Lemaitre, J.C.}, \bibinfo{author}{Azman, A.S.}, \bibinfo{author}{Luquero, F.J.}, \bibinfo{author}{Bertuzzo, E.}, \bibinfo{author}{Rinaldo, A.}, \bibinfo{year}{2018}.
\newblock \bibinfo{title}{Near real-time forecasting for cholera decision making in haiti after hurricane matthew}.
\newblock \bibinfo{journal}{PLoS Computational Biology} \bibinfo{volume}{14}, \bibinfo{pages}{e1006127}.
\bibitem[{Posny et~al.(2015)Posny, Wang, Mukandavire and Modnak}]{POSNY201538}
\bibinfo{author}{Posny, D.}, \bibinfo{author}{Wang, J.}, \bibinfo{author}{Mukandavire, Z.}, \bibinfo{author}{Modnak, C.}, \bibinfo{year}{2015}.
\newblock \bibinfo{title}{Analyzing transmission dynamics of cholera with public health interventions}.
\newblock \bibinfo{journal}{Mathematical Biosciences} \bibinfo{volume}{264}, \bibinfo{pages}{38--53}.
\bibitem[{Qian et~al.(2025)Qian, Marty, Basu, O'Dea, Wang, Fox, Rohani, Drake and Li}]{osti1706217}
\bibinfo{author}{Qian, Y.}, \bibinfo{author}{Marty, {\'E}.}, \bibinfo{author}{Basu, A.}, \bibinfo{author}{O'Dea, E.B.}, \bibinfo{author}{Wang, X.}, \bibinfo{author}{Fox, S.}, \bibinfo{author}{Rohani, P.}, \bibinfo{author}{Drake, J.M.}, \bibinfo{author}{Li, H.}, \bibinfo{year}{2025}.
\newblock \bibinfo{title}{Physics-informed deep learning for infectious disease forecasting}.
\newblock \bibinfo{journal}{arXiv preprint arXiv:2501.09298} .
\bibitem[{Rodr{\'\i}guez et~al.(2023)Rodr{\'\i}guez, Cui, Ramakrishnan, Adhikari and Prakash}]{rodriguez2023einns}
\bibinfo{author}{Rodr{\'\i}guez, A.}, \bibinfo{author}{Cui, J.}, \bibinfo{author}{Ramakrishnan, N.}, \bibinfo{author}{Adhikari, B.}, \bibinfo{author}{Prakash, B.A.}, \bibinfo{year}{2023}.
\newblock \bibinfo{title}{Einns: epidemiologically-informed neural networks}, in: \bibinfo{booktitle}{Proceedings of the AAAI Conference on Artificial Intelligence}, pp. \bibinfo{pages}{14453--14460}.
\bibitem[{Sardar et~al.(2013)Sardar, Mukhopadhyay, Bhowmick and Chattopadhyay}]{101371}
\bibinfo{author}{Sardar, T.}, \bibinfo{author}{Mukhopadhyay, S.}, \bibinfo{author}{Bhowmick, A.R.}, \bibinfo{author}{Chattopadhyay, J.}, \bibinfo{year}{2013}.
\newblock \bibinfo{title}{An optimal cost effectiveness study on zimbabwe cholera seasonal data from 2008–2011}.
\newblock \bibinfo{journal}{PLOS one} \bibinfo{volume}{8}, \bibinfo{pages}{1--18}.
\bibitem[{Sun et~al.(2017)Sun, Xie, Huang, Jin, Li and Liu}]{SUN2017235}
\bibinfo{author}{Sun, G.Q.}, \bibinfo{author}{Xie, J.H.}, \bibinfo{author}{Huang, S.H.}, \bibinfo{author}{Jin, Z.}, \bibinfo{author}{Li, M.T.}, \bibinfo{author}{Liu, L.}, \bibinfo{year}{2017}.
\newblock \bibinfo{title}{Transmission dynamics of cholera: Mathematical modeling and control strategies}.
\newblock \bibinfo{journal}{Communications in Nonlinear Science and Numerical Simulation} \bibinfo{volume}{45}, \bibinfo{pages}{235--244}.
\bibitem[{Trevisin et~al.(2022)Trevisin, Lemaitre, Mari and Pasetto}]{10982021}
\bibinfo{author}{Trevisin, C.}, \bibinfo{author}{Lemaitre, J.C.}, \bibinfo{author}{Mari, L.}, \bibinfo{author}{Pasetto, D.}, \bibinfo{year}{2022}.
\newblock \bibinfo{title}{Epidemicity of cholera spread and the fate of infection control measures}.
\newblock \bibinfo{journal}{Journal of The Royal Society Interface} \bibinfo{volume}{19}, \bibinfo{pages}{20210844}.
\bibitem[{Ye et~al.(2025)Ye, Pandey, Bawden, Sumsuzzman, Rajput, Shoukat, Singer, Moghadas and Galvani}]{Ye2025}
\bibinfo{author}{Ye, Y.}, \bibinfo{author}{Pandey, A.}, \bibinfo{author}{Bawden, C.}, \bibinfo{author}{Sumsuzzman, D.M.}, \bibinfo{author}{Rajput, R.}, \bibinfo{author}{Shoukat, A.}, \bibinfo{author}{Singer, B.H.}, \bibinfo{author}{Moghadas, S.M.}, \bibinfo{author}{Galvani, A.P.}, \bibinfo{year}{2025}.
\newblock \bibinfo{title}{Integrating artificial intelligence with mechanistic epidemiological modeling: a scoping review of opportunities and challenges}.
\newblock \bibinfo{journal}{Nature Communications} \bibinfo{volume}{16}, \bibinfo{pages}{581}.

\end{thebibliography}
\end{document}